\documentclass[paper]{JHEP}
\usepackage{psfig}
\usepackage{epsf}
\newcommand{\beq}{\begin{equation}}
\newcommand{\eeq}{\end{equation}}
\newcommand{\beqn}{\begin{eqnarray}}
\newcommand{\eeqn}{\end{eqnarray}}
\newcommand{\de}{{\mathrm{d}}}
\title{The decay $\tau\rightarrow 3 \pi \nu_{\tau}$ as a probe of the
mechanism of dynamical chiral symmetry breaking}
\author{Luca Girlanda and Jan Stern\\Division de Physique Th\'eorique, Institut de
Physique Nucl\'eaire\\ F-91406 Orsay Cedex\\ France}
\preprint{IPNO-TH/99-09}
\abstract{
The decays $\tau \rightarrow 3 \pi + \nu_{\tau}$ are analyzed at one loop
order in the framework of Generalized Chiral Perturbation Theory, in order
to test the sensitivity to the size of spontaneous chiral symmetry breaking
parameters, contained  in the S-wave.
The latter, due to a kinematical suppression, at threshold, of the P-wave,
is relatively large enough to be detectable at high energy machines, through
azimuthal left-right asymmetries.
This quantity (for the $\pi^-\pi^-\pi^+$ mode), integrated from
threshold to $Q^2 = 0.35$~GeV$^2$, varies from $(17 \pm 3 ) \%$  in the
standard case of large condensate up to $(40 \pm 5 )\%$ in the extreme case
of tiny condensate.
The feasibility of such measurement at high luminosity colliders ({\em e.g.}
CLEO) is discussed. This method provides a completely independent
cross-check of forthcoming experimental determination of the quark
condensate, based on low energy $\pi\pi$ scattering.
}
\begin{document}
\section{Introduction} \label{sec:intro}
The mechanism of dynamical (chiral) symmetry breaking and the identification
of its experimental signatures is one of the central
subjects in modern particle theory. The problem is to understand
non-perturbative
alternatives to the spontaneous symmetry breaking triggered by elementary
weakly coupled scalar fields, whose nature is essentially perturbative.
The problem arises at different energy scales characteristic of the Standard
Model and its extensions. At low energy QCD,  $E<1$~GeV, the nonperturbative
dynamical breaking
of chiral symmetry (DBCHS) is unavoidable, since QCD does not contain
elementary
weakly coupled scalars. Yet, the mechanism of DBCHS, in particular the
role and importance of the quark-antiquark condensate $\langle \bar q q
\rangle $, are not
understood theoretically and they remain poorely known experimentally
\cite{sternmainz}.
At higher energy scales $E \sim (10^2 - 10^3)$~GeV, a dynamical electroweak
symmetry breaking is of interest as a possible alternative to an elementary
weakly coupled light scalar Higgs protected by Supersymmetry.
This alternative assumes the existence of a new QCD-like (confining) gauge
theory which breaks dynamically its chiral symmetry at a scale
$\Lambda_{\mathrm{sb}} \sim (2-3)$~TeV. The
Goldstone bosons would then transmit this
DBCHS to the electroweak sector by the Higgs mechanism (for a recent review
see Ref.~\cite{chivu}) . In order to make
such a scenario specific and to investigate its consistency with electroweak
precision measurements at low energies ($E\ll\Lambda_{\mathrm{sb}}$), one
usually takes
the presumed mechanism of DBCHS in QCD as a prototype, assuming in addition, a
simple dependence on the scale and on the number of (techni-)colors
$N_{\mathrm{c}}$ and of
light (techni-)fermion dublets $N_{\mathrm{D}}$ \cite{pde}.
If the mechanism  of DBCHS in QCD-like theories turned out to be
 different from the general expectation  (concerning {\em e.g.} the role
of chiral condensates in a dynamical generation of fermion masses, the
possible existence of new phase transitions as a function of
$N_{\mathrm{c}}$ and $N_{\mathrm{D}}$ and
their influence on the spectrum of bound states, \ldots ), the estimates
of oblique and vertex corrections to the tree level Standard Model
\cite{pde} could be substantially modified\footnote{
A remark in this sense has been recently made in Ref.~\cite{eder},
concerning the value and the sign of the parameter $L_{10}$.}.
Below the scale $\Lambda_{\mathrm{sb}}$, at which new degrees of freedom
(``techni-hadrons'')  start to be visible,
the precision tests of models of dynamical electroweak symmetry breaking
can hardly be dissociated from the investigation of the mechanism of
DBCHS in low energy QCD. The importance of dedicated experimental tests of the
latter for the low-energy QCD phenomenology as it appears {\em e.g.} in the
calculation of weak matrix elements, determination of CKM matrix elements
or estimates of running light quark masses, has been often mentioned. (It
has, for instance, been pointed out \cite{marceuro} that if the condensate
$\langle \bar q q \rangle$ does
not dominate the expansion of symmetry breaking effects, the SM evaluation
of $\epsilon^{\prime}/\epsilon$ could lead to a significantly higher value
than  in the case
of the standard scenario of DBCHS). Actually,  model independent tests of the
role of $\langle \bar q q \rangle$ in QCD are of more fundamental and larger
interest: 
they may concern the mechanism of dynamical electroweak symmetry breaking at
the TeV scale and the problem of the origin of masses.

The purpose of this paper is to propose a new test of the importance of the
quark condensate in the mechanism of chiral symmetry breaking in QCD. The
experimental signature of this mechanism  can be systematically investigated
using  the
technique of the effective theory \cite{weinberg}, and its low energy
expansion - $\chi$PT \cite{gale1,gale2}. The standard version of this
expansion (S$\chi$PT) assumes a dominant role of $\langle \bar q q \rangle$
that is a 
small deviation from the Gell-Mann--Oakes--Renner (GOR) relation \cite{gor}. If
the condensate is less important the standard low energy expansion of the same
effective Lagrangian has to be modified leading to the Generalized $\chi$PT
(G$\chi$PT).
The latter is the proper framework for measuring $\langle \bar q q \rangle$
since it makes no assumptions on its size nor about the deviation from the
GOR relation\footnote{
The quark condensate by itself is not an observable quantity. What can be
actually measured is the renormalization group invariant product
$\hat m \langle \bar u u + \bar d d \rangle_{\hat m=0}$ in units of
$F_{\pi}^2 M_{\pi}^2$, that is precisely the deviation from the GOR
relation.}. 

In this way the low energy $\pi\pi$ scattering in the S-wave has been
identified as being particularly sensitive to $\langle \bar q q \rangle$
\cite{ssf,pipi1,marcref13}.
The corresponding experimental tests now become possible, due to $i)$~the
control of the precision of 
$\chi$PT which now reaches two-loop accuracy and is able to produce a
precise prediction of the standard scenario \cite{gale1,loropipi} and on the
other hand to  interpret any substantial deviation from this prediction in
terms of unexpectedely weak values of quark condensate \cite{pipi1};
$ii)$~the fact that the needed precision (5~\% or better for the S-wave
scattering length) can be reached: in the on-going $\pi^+-\pi^-$ atoms
lifetime measurement at
CERN \cite{dirac}, in the new precise $K^+_{e4}$ experiments, E865 at BNL
\cite{pislak} and, in the near future, at the  Frascati $\phi$ factory
Da$\phi$ne (with the KLOE detector) \cite{kloe}.

In view of the importance of this issue it is interesting to identify
independent sources of similar experimental information. In this paper a
proposal is elaborated in detail concerning the decays $\tau \rightarrow 3
\pi+\nu_{\tau}$ at low invariant mass of the hadronic system $Q^2 <
0.35$~GeV$^2$.
The S-wave is proportional to the divergence of the axial current and it
measures  the importance of chiral symmetry breaking away from the pion pole. 
However, when considering this process in $\chi$PT, one is faced with two
main difficulties: $i)$~the smallness of the branching ratio in the
threshold region, which is the only domain where $\chi$PT is applicable, and
$ii)$~ the chiral suppression of the S-wave compared to the P-wave
(which, moreover is enhanced by the  $a_1$ resonance).
The former problem now becomes less stringent, as pointed out in
Ref.~\cite{colangelo}, thanks to the large statistics already accumulated in
the present machines and to the improvements expected for the near future
(for a review see Ref.~\cite{perl}).
As for the latter, it turns out (cfr. Ref.~\cite{km}) that the P-wave,
although dynamically dominant, is kinematically suppressed, near the
threshold, compared to the S-wave.
These two facts together allow the S-wave contribution to show up in a clean
and detectable way.
The plan of the paper is as follows.
In Section~\ref{sec:lagr} the generalized SU(2)$\times$SU(2) chiral
lagrangian is 
constructed and renormalized at ${\mathcal O}(p^4)$ level. In
Section~\ref{sec:kine} we will
recall the kinematics of the $\tau\rightarrow 3 \pi \nu_{\tau}$ decays and
the definitions of the structure functions, following Ref.~\cite{km}. In
Section~\ref{sec:matr} we compute the hadronic matrix element at one-loop of
G$\chi$PT, and in Section~\ref{sec:pipi} we discuss the relation with
$\pi\pi$ observables.
The numerical results for the structure functions and their dependence on
the quark condensate are shown in Section~\ref{sec:num}. In particular we
will focus on 
the azimuthal asymmetries, which are very straightforward observables to
extract experimentally. In Section~\ref{sec:qmas} we discuss the relevance
of our results for the QCD sum rules used to extract the quark masses.
Finally we will summarize with some conlcuding remarks in
Section~\ref{sec:conc}. 
\section{The generalized expansion of the SU(2)$\times$SU(2) lagrangian}
\label{sec:lagr}
$\chi$PT is an effective theory having as degrees of freedom only the lowest
energy excitations of the spectrum of QCD, that is the Goldstone bosons. In
the case of SU(2)$\times$SU(2) chiral symmetry these are the pion fields
$\pi^j(x)$.
They are collected in a unitary $2\times2$ matrix
\beq
U(x)=\exp\left[\frac{ i \pi^j(x)\tau^j}{F}\right], \hspace{1cm} j=1,2,3,
\eeq
with $\tau^j$ the Pauli matrices, and $F$ the pion decay constant in the
chiral limit.
The external sources coupled to the vector, axial-vector, scalar and
pseudoscalar quark currents are also represented by hermitian $2\times2$
matrices
\beq
v_{\mu} = \sum_{j=0}^3 v_{\mu}^j \tau^j,\hspace{1cm} a_{\mu} = \sum_{j=1}^3
a_{\mu}^j 
\tau^j,\hspace{1cm}  s=\sum_{j=0}^3 s^j \tau^j, \hspace{1cm} p= \sum_{j=0}^3
p^j\tau^j. 
\eeq
 The scalar source (which contains the quark mass matrix) and the
 pseudoscalar one are grouped in a complex matrix $\chi =s+ i p$.  
The effective lagrangian is the most general one which is invariant under
chiral symmetry 
and it contains an infinite number of terms; however, only a finite number of
them will contribute at a given order of the low energy expansion.
Since the latter is a simultaneous expansion in powers of momenta and quark
masses, one has to establish a power counting rule for the quark masses.
This can be done by considering the Taylor expansion of the pion
mass squared in powers of quark mass (we will
always stick to the isospin limit, $m_{\mathrm{u}}=m_{\mathrm{d}}=\hat m$)
\beq \label{eq:mpi}
M_{\pi}^2= 2 B \hat m + 4 A \hat m^2 + \ldots ,
\eeq
which is valid up to chiral logarithms.
The coefficient of the linear term is proportional to the quark condensate,
\beq
B= - \lim_{\hat m\to 0} \frac{\langle \bar u u \rangle}{F^2} = - \lim_{\hat m \to 0}
\frac{\langle \bar d d \rangle}{F^2}.
\eeq
The standard assumption is  that the latter is large enough for the first
term to
dominate the series~(\ref{eq:mpi}), yielding the standard counting rule
$\hat m \sim {\cal O}(p^2)$. In general, however, nothing prevents $B$ to be
much smaller than  one  may wish; in this case the counting rule has to be
modified and, to be consistent, one has to 
consider formally $B$ as a small parameter (notice that the parameter $B$
has been removed from the definition of the matrix $\chi$, compared to the
standard notation). The generalized counting is then defined by
\beq
B\sim\hat m\sim \chi \sim {\cal O}(p),
\eeq
so that the expansion of the effective lagrangian in G$\chi$PT reads
\beq
{\cal L}^{\mathrm{eff}}=\tilde {\cal L}^{(2)} + \tilde {\cal L}^{(3)} +
\tilde {\cal L}^{(4)} +
\ldots
\eeq
with
\beq
\tilde {\cal L}^{(d)}=\sum_{k+l+n=d}B^n {\cal L}_{(k,l)}
\eeq
and ${\cal L}_{(k,l)}$ containing $k$ derivatives and $l$ powers of quark
masses.
Defining the covariant derivative as usual
\beq
D_{\mu} = \partial_{\mu} U -i \left[ v_{\mu},U\right] - i
\left\{a_{\mu},U\right\},
\eeq
and denoting by $\langle \ldots \rangle$ the trace of the flavor $2\times2$
matrices, the leading, order ${\cal O}(p^2)$, lagrangian reads
\beqn
\tilde {\cal L}^{(2)} &=&
\frac{F^2}{4} \left\{ \langle D_{\mu} U^{\dagger} D^{\mu} U \rangle + 2 B
\langle U^{\dagger} \chi + \chi^{\dagger} U \rangle + A \langle \left(
U^{\dagger} \chi \right)^2 + \left( \chi^{\dagger} U \right)^2 \rangle
\right. \nonumber \\
&& \left. + Z^P \langle U^{\dagger} \chi - \chi^{\dagger} U\rangle^2 + h_0
\langle \chi^{\dagger} \chi \rangle + h_0^{\prime} \left( \det \chi + \det
\chi^{\dagger} \right) \right\}.
\eeqn
Terms which would be of order ${\cal O}(p^4)$ in the standard counting are
already present at the lowest order in the generalized case, reflecting the
fact that the linear and the quadratic contribution in Eq.~(\ref{eq:mpi})
can compete. Up to a given chiral order the S$\chi$PT
lagrangian is a subset of the G$\chi$PT one.
The last two terms are high energy counter terms, needed for
the renormalization of the theory, but they will not appear in the expressions
of matrix elements of pions on the mass shell.
Using the equations of motion, field redefinitions, the
identity
\beq
M^2 - M \langle M \rangle + \det M=0
\eeq
valid for any $2\times 2$ matrix $M$, and the fact that
\beq
D_{\mu}U^{\dagger} D^{\mu} U = \frac{1}{2} \langle D_{\mu}U^{\dagger} D^{\mu} 
U \rangle,
\eeq
to absorb the redundant terms, we arrive at the following expressions for
the $\tilde {\cal L}^{(3)}$ lagrangian\footnote{
The same  lagrangian, up to  and including ${\cal O}(p^4)$, has been found,
independently, by Marc Knecht \cite{marc}, to whom we  are indebted for useful
correspondence. Notice that in this reference different notations have been
chosen for the pure source terms $h_0^{\prime}$ and $h_{(2,2)}$.} 
\beqn
\tilde {\cal L}^{(3)} &=& \frac{1}{4} F^2 \left\{ \xi^{(2)} \langle D_{\mu}
U^{\dagger} D^{\mu} U \left( \chi^{\dagger} U + U^{\dagger} \chi \right)
\rangle + \rho_1^{(2)} \langle \left( \chi^{\dagger} U \right)^3 + \left(
U^{\dagger} \chi \right)^3 \rangle \right. \nonumber \\
&& \left. + \rho_2^{(2)} \langle \left( \chi^{\dagger} U + U^{\dagger} \chi
\right) \chi^{\dagger} \chi \rangle + \rho_3^{(2)} \langle \chi^{\dagger} U
- U^{\dagger} \chi \rangle \langle \left( \chi^{\dagger} U \right)^2 -
\left( U^{\dagger} \chi \right)^2 \rangle \right. \nonumber \\
&& \left. + \rho_4^{(2)} \langle \chi^{\dagger} U
+ U^{\dagger} \chi \rangle \langle \left( \chi^{\dagger} U \right)^2 +
\left( U^{\dagger} \chi \right)^2 \rangle + \rho_5^{(2)}  \langle
\chi^{\dagger} \chi \rangle \langle \chi^{\dagger} U + U^{\dagger} \chi
\rangle \right\},
\eeqn
and for  $\tilde {\cal L}^{(4)} = {\cal L}_{(4,0)} + {\cal L}_{(2,2)} + {\cal
L}_{(0,4)}$,
\beqn
{\cal L}_{(4,0)} &=& \frac{l_1}{4} \langle D_{\mu} U^{\dagger}
D^{\mu} U \rangle^2 + \frac{l_2}{4} \langle D_{\mu} U^{\dagger} D_{\nu} U
\rangle \langle D^{\mu} U^{\dagger} D^{\nu} U \rangle + l_5 \langle
F_{\mu\nu}^R U F^{L,\mu\nu} U^{\dagger} \rangle   \nonumber \\
&& + i \frac{l_6}{2} \langle F_{\mu\nu}^R D^{\mu} U D^{\nu} U^{\dagger} +
F_{\mu\nu}^L D^{\mu}U^{\dagger} D^{\nu} U \rangle \nonumber \\
&& -\left(2  h_2 +
\frac{1}{2} l_5 \right) \langle F_{\mu\nu}^R
F^{R,\mu\nu} + F{\mu\nu}^L F^{L,\mu\nu} \rangle . \\
&& \nonumber \\
{\cal L}_{(2,2)} &=& \frac{1}{4} F^2 \left\{ a_1 \langle
D_{\mu}U^{\dagger} D^{\mu} U \left( \chi^{\dagger} \chi + U^{\dagger} \chi
\chi^{\dagger} U \right) \rangle + a_2 \langle D_{\mu} U^{\dagger} U
\chi^{\dagger} D^{\mu} U U^{\dagger} \chi \rangle  \right. \nonumber \\
&& \left. + a_3 \langle D_{\mu} U^{\dagger} U \left( \chi^{\dagger} D^{\mu}
\chi - D^{\mu} \chi^{\dagger} \chi \right) + D_{\mu} U U^{\dagger} \left(
\chi D^{\mu} \chi^{\dagger} - D^{\mu} \chi \chi^{\dagger} \right) \rangle
\right. \nonumber \\
&&\left. + b_1 \langle D_{\mu} U^{\dagger} D^{\mu} U \left( \chi^{\dagger} U
\chi^{\dagger} U + U^{\dagger} \chi U^{\dagger} \chi \right) \rangle \right.
\nonumber \\
&& \left. + b_2
\langle D_{\mu} U^{\dagger} \chi D^{\mu} U^{\dagger} \chi + \chi^{\dagger}
D_{\mu} U \chi^{\dagger} D^{\mu} U \rangle  \right. \nonumber \\
&&\left. + b_3 \langle U^{\dagger} D_{\mu} \chi U^{\dagger} D^{\mu} \chi +
D_{\mu} \chi^{\dagger} U D^{\mu} \chi^{\dagger} U \rangle \right. \nonumber
\\
&& \left. + c_1 \langle
D_{\mu} U^{\dagger} \chi + \chi^{\dagger} D_{\mu} U \rangle \langle D^{\mu}
U^{\dagger} \chi + \chi^{\dagger} D^{\mu} U \rangle  \right. \nonumber \\
&& \left. + c_2 \langle D_{\mu}\chi^{\dagger} U + U^{\dagger} D_{\mu} \chi
\rangle \langle D^{\mu} U^{\dagger} \chi + \chi^{\dagger} D^{\mu} U \rangle
\right. \nonumber \\
&& \left.
+ c_3 \langle D_{\mu} \chi^{\dagger} U + U^{\dagger} D_{\mu} \chi \rangle
\langle D^{\mu} \chi^{\dagger} U + U^{\dagger} D^{\mu} \chi \rangle \right.
\nonumber \\
&&\left. +  c_4 \langle D_{\mu} U^{\dagger} \chi - \chi^{\dagger} D_{\mu} U
\rangle \langle D^{\mu} U^{\dagger} \chi - \chi^{\dagger} D^{\mu} U \rangle
\right. \nonumber \\
&&\left.
+ c_5 \langle D_{\mu} \chi^{\dagger} U - U^{\dagger} D_{\mu} \chi \rangle
\langle D^{\mu} \chi^{\dagger} U - U^{\dagger} D^{\mu} \chi \rangle  +
h_{(2,2)} \langle D_{\mu} \chi^{\dagger} D^{\mu} \chi \rangle \right\}. \\
&& \nonumber \\
{\cal L}_{(0,4)} &=& \frac{1}{4} F^2 \left\{ e_1 \langle \left(
\chi^{\dagger} U \right)^4 + \left( U^{\dagger} \chi \right)^4 \rangle + e_2
\langle \chi^{\dagger} \chi \left( \chi^{\dagger} U \chi^{\dagger} U +
U^{\dagger} \chi U^{\dagger} \chi \right) \rangle  \right. \nonumber \\
&& \left. + e_3 \langle \chi^{\dagger} \chi U^{\dagger} \chi \chi^{\dagger}
U \rangle + f_1 \langle \left( \chi^{\dagger} U \right)^2 + \left(
U^{\dagger} \chi \right)^2 \rangle^2  \right. \nonumber \\
&& \left. + f_2 \langle \left( \chi^{\dagger} U \right)^3 + \left(
U^{\dagger} \chi \right)^3 \rangle \langle \chi^{\dagger} U + U^{\dagger}
\chi \rangle + f_3 \langle \chi^{\dagger} \chi \left( \chi^{\dagger} U +
U^{\dagger} \chi \right) \rangle \langle \chi^{\dagger} U + U^{\dagger}
\chi \rangle \right. \nonumber \\
&&\left. + f_4 \langle \left( \chi^{\dagger} U \right)^2 + \left(
U^{\dagger} \chi \right)^2 \rangle \langle \chi^{\dagger} U + U^{\dagger}
\chi \rangle^2  + f_5  \langle \left( \chi^{\dagger} U \right)^3 - \left(
U^{\dagger} \chi \right)^3 \rangle \langle \chi^{\dagger} U - U^{\dagger}
\chi \rangle \right. \nonumber \\
&&\left. + h_4 \langle \chi^{\dagger} \chi \chi^{\dagger} \chi \rangle + h_5
\langle \chi^{\dagger} \chi \rangle \left(\det \chi + \det \chi^{\dagger}
\right) \right. \nonumber \\
&&\left. +h_6 \left( \det \chi + \det \chi^{\dagger} \right)^2 + h_7 \left(
\det \chi - \det \chi^{\dagger} \right)^2 \right\} .
\end{eqnarray}
The lagrangian ${\cal L}_{(4,0)}$ is the same as in Ref.~\cite{gale1}.
The remaining terms of the ${\cal O}(p^4)$ lagrangian of Ref.~\cite{gale1},
related to the explicit chiral symmetry breaking sector of the theory, have
been redefined; the correspondence reads:
\beq \label{eq:transl}
A= \frac{2 B^2}{F^2} (l_3 + l_4), \hspace{1cm} Z^P= \frac{B^2}{F^2} (-l_3 -
l_4 + l_7), \hspace{1cm} \xi^{(2)}=\frac{2 B}{F^2} l_4,
\eeq
\beq
h_0=\frac{4 B^2}{F^2} (-l_4 + h_1 + h_3), \hspace{1cm} h_0^{\prime} =
\frac{4 B^2}{F^2} (l_3 + h_1 - h_3).
\eeq
The divergences at ${\cal O}(p^4)$ in dimensional regularization are
absorbed by the following renormalization of the coupling
constants\footnote{ 
A more detailed description of the generating functional, as well as the
correspondence between the SU(2) and the SU(3) constants, will be given
elsewhere \cite{io}.}
\beq
\mathrm{const} = \mathrm{const}^r + \Gamma_{\mathrm{const}} \lambda,
\eeq
with
\beq
\lambda= \frac{1}{16 \pi^2} \mu^{d-4} \left\{ \frac{1}{d-4} -\frac{1}{2}
\left( \log 4 \pi + \Gamma^{\prime}(1) + 1 \right) \right\},
\eeq
and the $\Gamma_{\mathrm{const}}$, calculated with the standard heat kernel
technique, are collected in the Table~\ref{tab:lambdas}.
\TABLE{
\begin{tabular}{||c|l||c|l||c|l||}
\hline
$ {\mathrm{const}}	$&$ F^2 \Gamma_{\mathrm{const}}
	$&  $ {\mathrm{const}} 	$&$ F^2 \Gamma_{\mathrm{const}}
	$&  $ {\mathrm{const}} 	$&$ F^2 \Gamma_{\mathrm{const}} $\\
	\hline
$ A 	$&$ 3 B^2			$&$ a_1	$&$ -2 Z^P		$&$
e_1	$&$ -2 \left( 2 A^2 + 10 A Z^P + 11 {Z^P}^2 \right) $\\
$ Z^P	$&$ -\frac{3}{2} B^2	$&$ a_2	$&$ -12 Z^P		$&$ e_2	$&$
-4 \left( A^2 + 3 A Z^P + 4 {Z^P}^2 \right) $\\
$ h_0	$&$0 			$&$ a_3	$&$ 0			$&
$ e_3	$&$ -4 \left( 3 A^2 + 12 A Z^P + 16 {Z^P}^2 \right) $\\
$ h_0^{\prime} $&$ 6 B^2	   	$&
$ b_1	$&$ 6 \left( A+Z^P\right)		$&
$ f_1	$&$ 3 \left( A^2 + 4 A Z^P + 5 {Z^P}^2 \right) $\\
\cline{1-2}
$ \xi^{(2)}    $&$ 4 B		$&
$ b_2	$&$ -2 \left(A+Z^P\right)		$&
$ f_2	$&$ 2 \left( A^2 + 5 A Z^P + 4 {Z^P}^2 \right) $\\
$ \rho_1^{(2)} $&$ -4 B \left(A + Z^P\right) $&
$ b_3	$&$ 0			$&
$ f_3	$&$ 4 \left( A^2 + 5 A Z^P + 6 {Z^P}^2 \right) $\\
$ \rho_2^{(2)} $&$ -4 B \left(A -3 Z^P\right)$&
$ c_1	$&$ 2 \left( A+ 2 Z^P\right)		$&
$ f_4	$&$ -6 \left( A Z^P + {Z^P}^2 \right) $\\
$ \rho_3^{(2)} $&$ 2 B \left(A +3 Z^P\right) $&
$ c_2	$&$ 0			$&
$ f_5	$&$ 2 \left( A^2 + 5 A Z^P + 4 {Z^P}^2 \right) $\\
$ \rho_4^{(2)} $&$ 2 B \left(3 A + Z^P\right)$&
$ c_3	$&$ 0			$&
$ h_4	$&$ 4 \left( A^2 + 2 A Z^P  + 2 {Z^P}^2 \right) $\\
$ \rho_5^{(2)} $&$ 4 B \left(A -2 Z^P\right) $&
$ c_4 	$&$ 2 A			$&
$ h_5	$&$ -4 \left( A^2 + 7 A Z^P + 8 {Z^P}^2 \right) $\\
\cline{1-2}
$ l_1	$&$ \frac{1}{3}		$&
$ c_5	$&$ 0			$&
$ h_6	$&$ 2 \left( 2 A^2 + 4 A Z^P + 3 {Z^P}^2 \right) $\\
$ l_2	$&$ \frac{2}{3}		$&
$ h_{(2,2)} $&$ 0			$&
$ h_7 	$&$ -14 {Z^P}^2			    $\\
\cline{3-6}
$ l_5	$&$ -\frac{1}{6}		$&
$	$&$			$&
$	$&$			$\\
$ l_6	$&$ - \frac{1}{3}		$&
$	$&$			$&
$	$&$			$\\
$ h_2	$&$ \frac{1}{12}		$&
$	$&$			$&
$	$&$			$\\
\hline
\end{tabular}
\caption{\it The $\beta$ function coefficients of the low energy constants.
The scale 
dependence is $\mu \, (\de / \de \mu)  {\mathrm{const}}^r = -1/\left( 16
\pi^2 \right)
\Gamma_{\mathrm{const}}$.}
\label{tab:lambdas}
}
We see from the table that, due to the chiral counting of $B$, the
divergences come all at order $p^4$, as they should.
\section{Kinematics} \label{sec:kine}
There are two different charge modes in the decay $\tau\rightarrow 3 \pi
\nu_\tau$, the $2\pi^0\pi^-$ mode and the all charged one, $2\pi^-\pi^+$. The
relevant hadronic matrix elements are
\beq
\begin{array}{c}
H^{00-}_{\mu}(p_1,p_2,p_3) = \langle
\pi^0(p_1)\pi^0(p_2)\pi^-(p_3)|A_{\mu}^-|0\rangle, \\
H^{--+}_{\mu}(p_1,p_2,p_3)= \langle
\pi^-(p_1)\pi^-(p_2)\pi^+(p_3)|A_{\mu}^-|0\rangle,
\end{array}\eeq
with $A^-_{\mu}=\bar u \gamma_{\mu}\gamma_5 d$.
Both these matrix elements can be simply expressed as follows:
\beqn
H^{--+}_{\mu}(p_1,p_2,p_3) &=& \sqrt{2} \left[ H^{+-0}_{\mu}(p_3,p_2,p_1) +
H^{+-0}_{\mu}(p_3,p_1,p_2) \right],  \label{eq:isospin1} \\ 
H^{00-}_{\mu}(p_1,p_2,p_3) &=& \frac{1}{2} \left[ H^{--+}_{\mu}(p_2,p_3,p_1)
+ H^{--+}_{\mu}(p_1,p_3,p_2) - H^{--+}_{\mu}(p_1,p_2,p_3) \right] \nonumber  \\
&=& \sqrt{2} H^{+-0}_{\mu}(p_1,p_2,p_3),   \label{eq:isospin2} 
\eeqn
with
\beq \label{eq:acca+-0}
H^{+-0}_{\mu}(p_1,p_2,p_3) = \langle
\pi^+(p_1)\pi^-(p_2)\pi^0(p_3)|A_{\mu}^0|0\rangle, \\
\eeq
and
\beq
A^0_{\mu}=\bar u \gamma_{\mu} \gamma_5 u - \bar d \gamma_{\mu} \gamma_5 d.
\eeq
$H^{+-0}_{\mu}$  contains directly as pole the invariant amplitude
$A(s|t,u)$ of the elastic $\pi\pi$ scattering. 
Both Eqs.~(\ref{eq:isospin1})-(\ref{eq:isospin2}) follow from isospin
and charge conjugation symmetry. 
The most general Lorentz structure of the matrix elements compatible with
G-parity invariance is
\beq
H_{\mu}^{\mathrm{hfs}}(p_1,p_2,p_3)=V_{1\mu}F_1^{\mathrm{hfs}}(p_1,p_2,p_3)
+ V_{2\mu} F_2^{\mathrm{hfs}}(p_1,p_2,p_3) + V_{4\mu}
F_4^{\mathrm{hfs}}(p_1,p_2,p_3),
\eeq
where hfs (hadronic final state) stands for $00-$ or $--+$ and
\beq
\begin{array}{l}
V_1^{\mu}=p_1^{\mu} - p_3^{\mu} -  \frac{Q(p_1-p_3)}{Q^2} Q^{\mu} ,\\
V_2^{\mu}=p_2^{\mu} - p_3^{\mu} -  \frac{Q(p_2-p_3)}{Q^2} Q^{\mu} , \\
V_4^{\mu}=p_1^{\mu} + p_2^{\mu} + p_3^{\mu} = Q^{\mu}.
\end{array}
\eeq
The G-parity violating contribution $i \epsilon^{\mu,\alpha,\beta,\gamma}
p_{1\alpha} p_{2\beta} p_{3\gamma} F_3(p_1,p_2,p_3)$  would be present if
$m_{\mathrm{u}} \neq m_{\mathrm{d}}$. However, such a term could only
originate from the WZW anomaly, which does not depend on quark masses.
Therefore it will start to contribute only at the loop level, that is, in
the case of the anomaly, at order ${\cal O}(p^6)$.
Notice that, in general, $m_{\mathrm{u}} - m_{\mathrm{d}}$ corrections are
expected to be of smaller relative importance in the case of small
condensate, since the ratio  $\left( m_{\mathrm{u}} - m_{\mathrm{d}} \right)
/\left( m_{\mathrm{u}} + m_{\mathrm{d}}\right)$ should decrease.
For both the $2\pi^-\pi^+$ and $2\pi^0\pi^-$ final states, Bose
symmetry requires
\beq
F_2^{\mathrm{hfs}}(p_1,p_2,p_3)=F^{\mathrm{hfs}}_1(p_2,p_1,p_3).
\eeq
$F_1$ ($F_2$) and $F_4$ correspond respectively to a final hadronic state of
total spin~1 and~0, as it is easily seen in the hadronic rest frame.
The differential decay rate is given by
\beq \label{eq:decayrate}
\de \Gamma (\tau\rightarrow 3 \pi \nu_\tau) = \frac{(2\pi)^4}{2 M_{\tau}}
|{\cal M} |^2 \de \Phi_4.
\eeq
${\cal M}$ is the matrix element of the electroweak interaction
\beq
{\cal M}= V_{\mathrm{ud}} \frac{G_{\mathrm{F}}}{\sqrt{2}}
L^{\mu}H_{\mu}^{\mathrm{hfs}},
\eeq
with
\beq
L_{\mu}=\bar u_{\nu_{\tau}} (p_{\nu_{\tau}},s_{\nu_{\tau}}) \gamma_\mu \gamma_5
u_{\tau}(p_\tau,s_\tau),
\eeq
and $\de\Phi_4$ is the invariant phase space of four particles
\beq
\de\Phi_4 = \frac{1}{256} \frac{\de\Omega_{\mathrm{h}}}{(2\pi)^{12}}
\frac{M_{\tau}^2 - 
Q^2}{M_{\tau}^2} \frac{\de Q^2}{Q^2} \de \alpha \,\de \gamma\, \de \cos
\beta \, \de s_1\, \de s_2\,.
\eeq
$\de \Omega_{\mathrm{h}}$ is the solid angle element of the hadronic system
in the $\tau$ rest
frame and can be written, after integration over the azimuthal direction, as
\beq
\de \Omega_{\mathrm{h}} = 2 \pi \left( \de \cos \theta \right),
\eeq
where $\theta$ is then the angle between the laboratory and the hadronic
system in the $\tau$ rest frame.
$\alpha$, $\beta$ and $\gamma$  are the
Euler angles which describe the orientation of the hadronic system in the
laboratory frame (for detailed definitions see Ref.~\cite{km}, which we
follow closely). The hadronic invariant variables are 
\beq
s_1 = (p_2 + p_3)^2, \hspace{1cm} s_2=(p_1 + p_3)^2, \hspace{1cm} s_3 = (p_1
+ p_2 )^2,
\eeq
with $s_1 + s_2 + s_3 = Q^2 + 3 M_{\pi}^2$.

As shown by Kuhn and Mirkes \cite{km}, the matrix element squared can be
written in 
terms of 9 independent leptonic and hadronic real structure functions $L_X$ 
and $W_X$
\beq
|{\cal M}|^2 =V_{\mathrm{ud}}^2 \, \frac{G_{\mathrm{F}}^2}{2} \sum_X
L_X W_X.
\eeq
All the angular dependence is contained in the functions $L_X$'s while
the $W_X$'s only depend on the hadronic invariant mass $Q^2$ and on the
Dalitz plot variables $s_1$ and $s_2$.
Four of the hadronic structure functions correspond to the square of the
spin~1 part of the hadronic matrix element,
\beq
\begin{array}{l}
W_A= (x_1^2 + x_3^2) |F_1|^2 + (x_2^2 + x_3^2) |F_2|^2 + 2 (x_1 x_2 - x_3^2)
{\mathrm{Re}} (F_1 F_2^{*}), \\
W_C= (x_1^2 - x_3^2) |F_1|^2 + (x_2^2 - x_3^2) |F_2|^2 + 2 (x_1 x_2 + x_3^2)
{\mathrm{Re}}  (F_1 F_2^{*}), \\
W_D=2\left[ x_1 x_3 | F_1|^2 - x_2 x_3 |F_2|^2 + x_3 ( x_2 - x_1)
{\mathrm{Re}}  (F_1 F_2^{*}) \right], \\
W_E = - 2 x_3 (x_1 + x_2) {\mathrm{Im}}  (F_1 F_2^{*}),
\end{array}
\eeq
one is the square of the spin~0 component
\beq
W_{SA} = Q^2 |F_4|^2,
\eeq
and the remaining ones are the interference between the spin~0 and spin~1
components
\beq
\begin{array}{l}
W_{SB} = 2 \sqrt{Q^2} \left[ x_1 {\mathrm{Re}} (F_1 F_4^{*}) + x_2
{\mathrm{Re}} (F_2 F_4^{*}) \right], \\
W_{SC} = - 2 \sqrt{Q^2} \left[ x_1 {\mathrm{Im}} (F_1 F_4^{*}) + x_2
{\mathrm{Im}} (F_2 F_4^{*}) \right], \\
W_{SD} = 2 \sqrt{Q^2} x_3 \left[ {\mathrm{Re}} (F_1 F_4^*) - {\mathrm{Re}}
(F_2 F_4^*) \right], \\
W_{SE} = -2 \sqrt{Q^2} x_3 \left[ {\mathrm{Im}} (F_1 F_4^*) - {\mathrm{Im}}
(F_2 F_4^*) \right].
\end{array}
\eeq
The $x_i$ are kinematical functions which are linear in the pion
three-momenta in the hadronic rest frame\footnote{
This system is oriented in such a way that the $x$-axis is along the
direction of $\vec p_3$ and all the pion fly in the $x$--$y$ plane.}, $x_1 =
V_1^x$, $x_2 = V_2^x$ and 
$x_3=V_1^y$: they vanish at the production threshold and can be expressed as
follows: 
\beq
\begin{array}{c}
x_1 x_2 - x_3^2= (Q^2 /2 - s_3 - M_{\pi}^2/2 ) + ( s_3 - s_1) ( s_3 -
s_2)/(4 Q^2), \\
x_1 + x_2 =-3/2 \sqrt{h_0},\\
x_3 = \sqrt{h},\\
- h_0 = 4 M_{\pi}^2 - ( 2 M_{\pi}^2  - s_1 - s_2)^2/Q^2,\\
h= \left[ s_1 s_2 s_3 - M_{\pi}^2 (Q^2 - M_{\pi}^2)^2 \right] /(h_0 Q^2).
\end{array}
\eeq
We can see, from the explicit expressions for the $W_X$'s, that the purely
spin~1 structure
functions are suppressed, at threshold, by two powers of the $x_i$'s, the
interferences of spin~1 and spin~0 by one power whereas the purely spin~0
structure function is not suppressed at all.  This fact, as already pointed
out in the introduction, allows us to have, at threshold, a much larger
sensitivity to the S-wave, and then to the size of chiral symmetry breaking
measured by the divergence of the axial current.
\section{The hadronic matrix element in G$\chi$PT} \label{sec:matr}
We have seen how the hadronic matrix elements for both the charge modes can
be expressed in terms of $H^{+-0}(p_1,p_2,p_3)$. It is then sufficient to
compute this quantity: from the residue of the pion
pole we will immediately read the $\pi\pi$ scattering amplitude. The
G$\chi$PT result at one loop level is
\begin{eqnarray} \label{eq:H+-0}
&& i H^{+-0}_{\mu} (p_1,p_2,p_3) =  \frac{-F_{\pi} Q_{\mu}}{Q^2 - M_{\pi}^2} \,
A(s_3|s_1,s_2)
\nonumber \\
&& + \frac{Q_{\mu}}{F_{\pi}} \biggl\{  1 + \xi^{(2)r} \hat m - 2 (\xi^{(2)}
\hat m)^2 + ( 4 a_2^r + 4 a_3
+ 4 b_1^r + 4 b_2^r + 8 c_1^r - 4 c_2 ) \hat m^2 \nonumber \\
&& + 2 \frac{ l_1^r}{F_{\pi}^2} ( s_3 - 2 M_{\pi}^2) - \frac{l_6^r}{F_{\pi}^2}
( s_1 + s_2 -2  M_{\pi}^2 ) + \frac{1}{F_{\pi}^2} \left( \frac{1}{2} s_3 + 8
A \hat m^2 \right) J^r(s_3)
\nonumber \\
&& + \frac{1}{F_{\pi}^2} \left[ s_1 M^r(s_1) + s_2 M^r(s_2) \right]
\nonumber \\
&& -\frac{1}{32 \pi^2 F_{\pi}^2 } \left[ -\frac{1}{3} M_{\pi}^2 + 8 A \hat m^2 +
\frac{1}{6} (s_1  + s_2) -\frac{1}{9} Q^2\right] \nonumber \\
&& -\frac{1}{32 \pi^2 F_{\pi}^2 } \left[ - M_{\pi}^2 + 8 A \hat m^2 +
\frac{1}{2} (s_1 + s_2) -\frac{1}{3}  Q^2 \right] \log \frac{M_{\pi}^2}{\mu^2}
\biggr\}
\nonumber \\
&&- 2 \frac{p_{3\mu}}{F_{\pi}} \biggl[ 1 + \xi^{(2)r} \hat m - 2 (\xi^{(2)} \hat m)^2 + ( 3
a_2^r + 4 a_3 + 4 b_1^r
+ 2 b_2^r + 4 c_1^r - 2 c_2 ) \hat m^2 \nonumber \\
&&
+ \frac{2 l_1^r}{F_{\pi}^2} (s_3 - 2 M_{\pi}^2)
-\frac{l_2^r}{2 F_{\pi}^2} ( s_1 + s_2 - 4 M_{\pi}^2 ) - \frac{
l_6^r}{F_{\pi}^2}
Q^2 - 2 \frac{M_{\pi}^2}{F_{\pi}^2} k_{\pi\pi} + \frac{1}{192 \pi^2
F_{\pi}^2} (  s_1 + s_2) \nonumber \\
&&
+ \frac{1}{F_{\pi}^2} \left( \frac{s_3}{2} + 8 A \hat m^2 \right) J^r(s_3)
+ \frac{1}{F_{\pi}^2} \left(-\frac{1}{4} M_{\pi}^2 + A \hat m^2 \right)
\left( J^r(s_1) + J^r(s_2)\right) \biggr] \nonumber \\
&&
+\frac{1}{F_{\pi}^3}  \left(p_1 - p_2 \right)_{\mu} \biggl[ l_2^r ( s_2 -
s_1) + \left( \frac{1}{3} s_2 - \frac{5}{6} M_{\pi}^2 - 2 A \hat m^2 \right)
J^r(s_2) \nonumber \\
&&
- \left(\frac{1}{3} s_1 - \frac{5}{6} M_{\pi}^2 - 2 A \hat m^2 \right)
J^r(s_1) + \frac{1}{288 \pi^2} ( s_2 - s_1) \biggr].
\end{eqnarray}
The scale dependence of the l.e.c.'s is absorbed by the corresponding
dependence  in the standard loop functions $J^r(s)$ and $M^r(s)$,
defined {\em e.g.} in Ref.~\cite{gale2}, and in the chiral logarithm 
\beq
k_{\pi\pi}=\frac{1}{32 \pi^2} \left( \log \frac{M_{\pi}^2}{\mu^2} + 1 \right),
\eeq
so that the total result is independent of the
$\chi$PT scale.
$A(s_3|s_1,s_2)$ is the (scale independent) off shell $\pi\pi$ scattering
amplitude, with $s_1+s_2+s_3=3 M_{\pi}^2 + Q^2$,
\begin{eqnarray} \label{eq:ampipi}
A(s_3|s_1,s_2) &=& \frac{\beta}{F_{\pi}^2} \left( s_3 - \frac{4}{3} M_{\pi}^2
\right) + \alpha \frac{M_{\pi}^2}{3 F_{\pi}^2} \nonumber \\
&& + \frac{1}{48 \pi^2 F_{\pi}^4}\left( \bar l_1 - \frac{4}{3} \right)
\left( s_3 - 2 M_{\pi}^2 \right)^2 \nonumber \\
&& + \frac{1}{48 \pi^2 F_{\pi}^4} \left( \bar l_2 - \frac{5}{6} \right) \left[ \left( s_1 - 2 M_{\pi}^2 \right)^2 + \left(
s_2 - 2 M_{\pi}^2 \right)^2 \right]  \nonumber \\
&& + \frac{1}{F_{\pi}^4} \biggl\{ \frac{1}{2} \left[ s_3^2 - M_{\pi}^4 + 32 A
\hat m^2 s_3 - 24 A \hat m^2 M_{\pi}^2 + 112 A^2 \hat m^4 \right] \bar J
(s_3)\nonumber \\
&&+ \left[ \left(M_{\pi}^2 + 4 A \hat m^2 - \frac{1}{2} s_1 \right)^2 +
\frac{1}{12} (s_3 - s_2) ( s_1 - 4 M_{\pi}^2 ) \right] \bar J (s_1)
\nonumber \\ 
 && + \left[ \left(M_{\pi}^2 + 4 A \hat m^2 - \frac{1}{2} s_2 \right)^2 +
\frac{1}{12} (s_3 - s_1) ( s_2 - 4 M_{\pi}^2 ) \right] \bar J (s_2) \biggr\} ,
\end{eqnarray}
Putting $Q^2 = M_{\pi}^2$ one recovers the $\pi\pi$ amplitude written in the
same form as in Ref.~\cite{pipi1} (dropping ${\cal O}(p^6)$ contributions). 
The expressions of the scale-independent parameters $\alpha$, $\beta$, $\bar
l_1$ and $\bar l_2$ in terms of the low energy constants are given by the
following equations:
\beqn
\frac{F_\pi^2}{F^2} M_{\pi}^2 &=& 2 B \hat m + 4 A^r \hat m^2
 + \left( 9 \rho_1^r + \rho_2^r + 20 \rho_4^r + 2 \rho_5^r \right) \hat
m^3  \nonumber \\
&& + \left( 16 e_1^r + 4 e_2^r + 32 f_1^r + 40 f_2^r  + 8 f_3^r + 96 f_4^r
\right) \hat m^4  \nonumber \\
&& + 4 a_3 M_{\pi}^2 \hat m^2 -
\frac{M_{\pi}^2}{32 \pi^2 F_{\pi}^2}  \left( 3 M_{\pi}^2 + 20 A \hat m^2
\right) \log \frac{M_{\pi}^2}{\mu^2}, \\
F_{\pi}^2 &=& F^2 \biggl[ 1 + 2 \xi^{(2),r} \hat m + \left( 2 a_1^r + a_2^r +
4 a_3 + 2 b_1^r - 2 b_2^r \right) \hat m^2  \nonumber \\
&& \left. - \frac{M_{\pi}^2}{8 \pi^2 F_{\pi}^2} \log \frac{M_{\pi}^2}{\mu^2}
\right], \\ 
\beta &=& 1 + 2 \xi^{(2),r} \hat m - 4 {\xi^{(2)}}^2 \hat m^2 + 2 \left( 3 a_2^r 
+ 2 a_3 + 4 b_1^r + 2 b_2^r + 4 c_1^r \right) \hat m^2  \nonumber  \\
&& - \frac{4 M_{\pi}^2}{32 \pi^2 F_{\pi}^2} \left( 1 + 10 \frac{ A \hat
m^2}{M_{\pi}^2 } \right) \left( \log \frac{M_{\pi}^2}{\mu^2} + 1 \right) ,
\label{eq:xibeta}  \\
\alpha M_{\pi}^2 \frac{F_{\pi}^2}{F^2} &=& 2 B \hat m + 16 A^r \hat m^2 -
4 M_{\pi}^2 \xi^{(2),r} \hat m \nonumber \\
&& + \left( 81 \rho_1^r + \rho^r_2 + 164 \rho_4^r + 2 \rho_5^r \right) \hat
m^3  \nonumber \\
&& - 8 M_{\pi}^2 \left( 2 b_1^r - 2 b_2^r - a_3 - 4 c_1^r \right) \hat m^2 
\nonumber \\
&& + 16 \left( 6 A a_3 + 16 e_1^r + e_2^r + 32 f_1^r + 34 f_2^r + 2 f_3^r +
72 f_4^r \right) \hat m^4  \nonumber \\
&& - \frac{M_{\pi}^2}{32 \pi^2 F_{\pi}^2} 
\left( 4 M_{\pi}^2 + 204 A \hat m^2  + 528 \frac{ A^2 \hat m^4}{M_{\pi}^2}
\right) \log \frac{M_{\pi}^2}{\mu^2}  \nonumber \\ 
&& - \frac{ 1}{32 \pi^2 F_{\pi}^2} \left[ M_{\pi}^4 + 88 A \hat m^2
M_{\pi}^2 + 528 A^2 \hat m^4 \right], \label{eq:Aalpha} \\
\bar l_1 &=& 96 \pi^2   l_1^r(\mu)  - \log \frac{M_{\pi}^2}{\mu^2}, \\
\bar l_2 &=& 48 \pi^2   l_2^r(\mu)  - \log \frac{M_{\pi}^2}{\mu^2}.
\eeqn
The parameter $\alpha$ contains the main information about the size of the
quark condensate: at tree level it varies
between~1 and~4 if $\langle \bar q q\rangle$ decreases from its standard
value down to zero\footnote{
The relation between $\langle \bar q q \rangle$ and $\alpha$ at one-loop
level will be revisited in another publication \cite{ioconstern}.}, while
$\beta$ stays always close to~1.
Solving Eq.~(\ref{eq:Aalpha}) for $A$ and Eq.~(\ref{eq:xibeta}) for
$\xi^{(2)}$, we can express the form factors in terms of $\alpha$ and
$\beta$.
The result for the $2\pi^0\pi^-$ mode in terms of the form factors
$F_1(p_1,p_2,p_3)$ and $F_4(p_1,p_2,p_3)$ introduced in
Section~\ref{sec:kine} is:
\beqn
\frac{F_{\pi}}{\sqrt{2}} F_1^{\mathrm{00-}}
&=&
\frac{1}{3} (\beta + 1)  + \frac{4}{3} ( a_3  - c_2) \hat m^2 + \frac{1}{96
\pi^2 F_{\pi}^2} \left[-4 M_{\pi}^2  - 2 Q^2  + \frac{2}{3} s_2 + 4 s_1 \right.
\nonumber \\
&&
\left. + \frac{4}{3} \bar l_1 ( s_3 - 2 M_{\pi}^2 ) + \frac{4}{3} \bar l_2 (
2 M_{\pi}^2 +  s_2 -2  s_1 ) + \frac{2}{3} \bar l_6 Q^2 \right]  \nonumber \\
&&
+ \frac{1}{3 F_{\pi}^2} \left[ s_3 + \frac{4}{3} M_{\pi}^2 ( \alpha -1)
\right] \bar J(s_3) + \frac{1}{3 F_{\pi}^2} \left[ s_2 - \frac{2}{3} M_{\pi}^2 (
\frac{\alpha}{2} + 4 ) \right] \bar J(s_2) \nonumber \\
&&
+ \frac{1}{3 F_{\pi}^2} \left[ - s_1 + \frac{2}{3} M_{\pi}^2 ( \alpha + 2
)\right] \bar J(s_1), \label {eq:f1}\\
&& \nonumber \\
\frac{F_{\pi}}{\sqrt{2}} F_4^{\mathrm{00-}}
&=&
 \frac{ s_3 - M_{\pi}^2}{2 Q^2} ( \beta +1 ) - \frac{1}{Q^2 - M_{\pi}^2} \left[
\beta \left( s_3 - \frac{4}{3} M_{\pi}^2 \right) + \frac{ \alpha }{3}
M_{\pi}^2\right] \nonumber \\
&&  + \left[ a_2^r + 2 a_3 + 2 b_2^r + 4 c_1^r - 4 c_2 - 2 \frac{s_2+s_1-2
M_{\pi}^2 }{Q^2} (a_3 - c_2) \right] \hat m^2 \nonumber \\
&& - \frac{M_{\pi}^2}{96 \pi^2
F_{\pi}^2} (\alpha -1) \log \frac{M_{\pi}^2}{\mu^2}  \nonumber \\
&&
+ \frac{1}{96 \pi^2 F_{\pi}^2} \left\{ M_{\pi}^2 \left[ 6  \frac{s_2 + s_1 - 2
M_{\pi}^2}{ Q^2} -  \alpha - 1 \right] + 3 ( s_2 + s_1 - 2 M_{\pi}^2 ) \right.
\nonumber \\
&&
- \frac{8}{3} Q^2  + \frac{ 7}{6} \frac{ s_2 + s_1}{Q^2} ( 2 Q^2 - 3 s_2 - 3 s_1 + 6
M_{\pi}^2 ) - \frac{5}{6} \frac{(s_2-s_1)^2}{Q^2} \nonumber \\
&&
 + 2 \bar l_1 \frac{ Q^2 - s_2 - s_1 + 2 M_{\pi}^2 }{Q^2} (s_3 - 2 M_{\pi}^2
)  \nonumber \\
&&
\left.  + \bar l_2 \frac{1}{Q^2} \left[ ( s_2 - s_1)^2 + ( s_2 + s_1 - 2 M_{\pi}^2
) ( s_2 + s_1 - 4 M_{\pi}^2 ) \right] \right\} \nonumber \\
&&
+\frac{1}{12 F_{\pi}^2} \left\{ s_2 - 4 M_{\pi}^2 + \frac{  s_2 + s_1 - 2
M_{\pi}^2 }{Q^2} M_{\pi}^2 (4 - \alpha  ) \right. \nonumber \\
&& + \frac{s_2 - s_1}{ Q^2} \left[ 2 s_2 -
M_{\pi}^2 ( 4 + \alpha) \right]  \nonumber \\
&&
\left. - \frac{1}{Q^2 - M_{\pi}^2} \left[ 3 \left( s_2 - \frac{2}{3} (\alpha +
2 ) M_{\pi}^2 \right)^2 + (s_3-s_1) ( s_2 - 4 M_{\pi}^2 ) \right] \right\} \bar
J(s_2) \nonumber \\ 
&&
+\frac{1}{12 F_{\pi}^2} \left\{ s_1 - 4 M_{\pi}^2 + \frac{  s_2 + s_1 - 2 M_{\pi}^2
}{Q^2} M_{\pi}^2 (4 - \alpha  ) \right. \nonumber \\
&& + \frac{s_1 - s_2}{ Q^2} \left[ 2 s_1 - M_{\pi}^2
( 4 + \alpha) \right]  \nonumber \\
&&
\left. - \frac{1}{Q^2 - M_{\pi}^2} \left[ 3 \left( s_1 - \frac{2}{3} (\alpha +
2 ) M_{\pi}^2 \right)^2 + (s_3-s_2) ( s_1 - 4 M_{\pi}^2 ) \right]
\right\} \bar J(s_1) \nonumber \\
&&
+ \frac{1}{2 F_{\pi}^2} \left\{ \frac{1}{Q^2} \left(  Q^2 - s_2 - s_1 + 2
M_{\pi}^2 \right) \left[ s_3 + \frac{4}{3} M_{\pi}^2 ( \alpha -1) \right]
\right. \nonumber \\
&&
\left. - \frac{1}{Q^2 - M_{\pi}^2} \left[ s_3^2 + \frac{8}{3} M_{\pi}^2 (\alpha -1 ) s_3
+ M_{\pi}^4 \left( \frac{7}{9} \alpha^2 - \frac{32}{9} \alpha + \frac{16}{9}
\right) \right] \right\} \bar J(s_3) \nonumber \\
&&
-\frac{F_{\pi}^2 }{Q^2 - M_{\pi}^2} \left\{ \frac{1}{48 \pi^2 F_{\pi}^4}
\left( \bar l_1 - \frac{4}{3} \right) (s_3 - 2
M_{\pi}^2 )^2 \right. \nonumber \\
&& \left. + \frac{1}{48 \pi^2 F_{\pi}^4} \left( \bar l_2 - \frac{5}{6}
\right) \left[ (s_2- 2 M_{\pi}^2 )^2 + ( s_1
- 2 M_{\pi}^2 )^2 \right] \right\},\label{eq:f4}
\end{eqnarray}
where we have defined the scale independent quantities
\beq
\bar l_i = \frac{32 \pi^2}{\gamma_i} l_i^r(\mu) - \log
\frac{M_{\pi}^2}{\mu^2},
\eeq
and $\gamma_i$ are the corresponding $\beta$ function coefficients
collected in Table~\ref{tab:lambdas}.
It is easy to verify, taking the S$\chi$PT expressions for $\alpha$ and
$\beta$,
\beqn 
\alpha^{\mathrm{st}} &=& 1 + \left[ -\frac{1}{32 \pi^2} \left( \log
\frac{M_{\pi}^2}{\mu^2} + 1 \right) + 6
l_3^r + 2 l_4^r \right] \frac{M_{\pi}^2}{F_{\pi}^2},
\label{eq:alphastandard} \\ 
\beta^{\mathrm{st}} &=& 1 + \left[ -\frac{1}{8\pi^2} \left( \log \frac{M_{\pi}^2}{\mu^2} +
1 \right) + 2 l_4^r \right] \frac{M_{\pi}^2}{F_{\pi}^2}.\label{eq:betastandard}
\eeqn
that one recovers the results of Ref.~\cite{colangelo}.
Notice that the form factor $F_4$ is proportional to the matrix element of
the divergence of the axial current. One easily verifies that the
expression~(\ref{eq:f4}) goes to zero in the chiral limit. These two facts,
 together with the global renormalization scale independence, provides
 non-trivial checks of our calculation.
It is remarkable that in Eqs.~(\ref{eq:f1}) and (\ref{eq:f4}) all the
contributions from the $\tilde {\cal L}^{(3)}$ lagrangian are absorbed in the
definitions of $\alpha$, $\beta$, $F_{\pi}$ and $M_{\pi}$.
\section{Connection with low-energy $\pi\pi$ observables} \label{sec:pipi}
The renormalization-group invariant parameters $\alpha$  and $\beta$ are
essentially the (on shell) $\pi\pi$
amplitude $A(s|t,u)$ and its slope at the symmetrical point $s=t=u=4/3
M_{\pi}^2$. 
For this reason one expects both $\alpha$ and $\beta$ to be less sensitive
to the higher order 
chiral corrections than, say, the S-wave scattering lengths.
They
represent the ${\cal O}(p^4)$ G$\chi$PT truncation of observable quantities
$\alpha$ and $\beta$ defined, including ${\cal O}(p^6)$ accuracy, in
Ref.~\cite{pipi1},
\begin{equation} \label{eq:ampl}
 A(s|t,u) = A_{\mathrm{KMSF}}(s|t,u;\alpha,\beta;\lambda_1, \lambda_2,
 \lambda_3, 
 \lambda_4) + {\cal O} \left[ \left({p \over \Lambda_{\mathrm{H}}} \right)^8
 , \left( {M_{\pi} \over 
 \Lambda_{\mathrm{H}}} \right)^8 \right],
 \end{equation}
where the function $A_{\mathrm{KMSF}}$ is expicitly displayed.
Both $\alpha$
and $\beta$  will be measured in forthcoming high-precision low-energy
$\pi\pi$ experiments (Dirac at CERN, E865 at BNL, KLOE at Da$\Phi$ne).
The values of $\alpha$ and $\beta$ are constrained by the behavior of the
phase $\delta_1^1$ in the $\rho$ region, using the Roy dispersion relations
\cite{roy} (resulting in the so-called Morgan-Shaw universal curve
\cite{morgan}).
The parameters $\lambda_1,\ldots \lambda_4$ have been determined in
Ref.~\cite{pipi2} from medium-energy experimental $\pi\pi$ phases, using a
set of rapidly converging sum rules derived from Roy dispersion relations.
The results depend very little on the size of $\langle \bar q q \rangle$.
Comparing the expression~(\ref{eq:ampl}) with the explicit two-loop
calculation performed in S$\chi$PT \cite{l3l4} one finds the relationship
between the parameters $\lambda_i$'s and the low energy
constants of the Standard lagrangian \cite{l1l2},
\beq \label{eq:laivsli}
\lambda_i = \lambda_i ( \bar l_1, \bar l_2, \bar l_4,\ldots ), \hspace{1cm}
i=1,\ldots 4.
\eeq
Notice that the parameters~(\ref{eq:laivsli}) do not depend at all on the
critical standard ${\cal O}(p^4)$ constant $\bar l_3$, which controls the
size of $\langle \bar q q \rangle$, and the dependence on the constant $\bar
l_4$ shows up only at ${\cal O}(p^6)$ order. For this reason
Eq.~(\ref{eq:laivsli})  can be inverted for $\bar l_1$ and $\bar l_2$,
yielding the values \cite{l1l2}
\beq \label{eq:lis}
\bar l_1 = -0.37 \pm 1.96, \hspace{1cm} \bar l_2 =4.17 \pm 0.47.
\eeq
where the error bars come from the uncertainty on the parameters
$\lambda_i$'s and the estimated uncertainty of higher chiral orders.
Strictly speaking the values of Eq.~(\ref{eq:lis}) have been obtained within
the framework of S$\chi$PT.
 However, for the reason mentioned above, the ${\cal
L}_{(4,0)}$ constants $\bar l_1$ and $\bar l_2$ should not be affected by the
variation of the quark condensate.
Improvements in the precision of the determination of $l_1$ and $l_2$ are
expected from forthcoming numerical solutions of Roy Equations
\cite{colaroy}. 

Using the currently available data on $\delta_0^0 - \delta_1^1$ extracted
from the last $K^+_{e4}$  experiment (Rosselet {\em et al.} \cite{rosselet})
the best fitted  values 
 for $\alpha$ and $\beta$ are \cite{pipi1}:
\beq  \label{eq:alfabetaexp}
\alpha^{\mathrm{exp}}=2.16\pm0.86, \hspace{1cm} \beta^{\mathrm{exp}}=1.074
\pm 0.053,
\eeq
where the error bars are merely experimental.
They have to be compared with the S$\chi$PT ${\cal O}(p^4)$
expressions~(\ref{eq:alphastandard})-(\ref{eq:betastandard}), which, using
for $l_3$ and $l_4$ the central values quoted in Ref.~\cite{l3l4}
\beq \label{eq:l3l4}
l_3^r(M_{\rho}) = 0.82 \times 10^{-3}, \hspace{1cm} l_4^r(M_{\rho}) = 5.6
\times 10^{-3},  
\eeq
give 
\beq
\alpha^{\mathrm{st}}=1.06, \hspace{1cm} \beta^{\mathrm{st}} = 1.095.
\eeq
Taking further into account the higher order contributions,  estimated in
Ref.~\cite{l1l2}, the standard values for $\alpha$ and $\beta$ become\footnote{
Both the values~(\ref{eq:lis}) and the S$\chi$PT
prediction~(\ref{eq:alfabetast}) are compatible with the central values
referred to as ``Set II'' given in Ref.~\cite{loropipi} without error bars.
Notice that the errorbars given in Eq.~(\ref{eq:alfabetast}) do not include
any uncertainty  due to $l_3$ and $l_4$. Eq.~(\ref{eq:l3l4}) is taken as our
{\em definition} of the standard case.}
\beq \label{eq:alfabetast}
\alpha^{\mathrm{st}}=1.07 \pm 0.01, \hspace{1cm} \beta^{\mathrm{st}} = 1.105
\pm 0.015.
\eeq
Our strategy to detect the sensitivity to the size of  $\langle \bar q
q\rangle$, cross-checking the future determinations from $\pi\pi$
scattering, will be to plot the structure functions for three sets of values
of  $\alpha$ and $\beta$, 
\beq \label{eq:trealfabeta}
\begin{array}{ll}
\alpha^{\mathrm{st}}=1.07, & \beta^{\mathrm{st}} = 1.105,\\
\alpha^{\mathrm{exp}}=2.16, & \beta^{\mathrm{exp}}=1.074, \\
\alpha = 4 , & \beta=1.16.
\end{array}
\eeq
They correspond respectively to the S$\chi$PT predictions,
Eq.~(\ref{eq:alfabetast}),  the  central values obtained from the best
fit to the Rosselet {\em et al.} data \cite{rosselet},
Eq.~(\ref{eq:alfabetaexp})  and the extreme 
case of tiny condensate ($\alpha=4$ with the corresponding $\beta=1.16$
inferred from the Morgan-Show universal curve). 

The other numerical input for the form factors (not entering the $\pi\pi$
amplitude) that we will use is the value of $\bar l_6$, from
the two-loop analysis of the vector form factor of the pion \cite{bicota},
\beq \label{eq:l6b}
\bar l_6 = 16.0 \pm 0.9 .
\eeq
The contributions from the low energy constants of ${\cal L}_{(2,2)}$ and
${\cal L}_{(0,4)}$ will be considered as sources of error, with zero central
values and errors according to na\"{\i}ve dimensional analysis estimates,
that is
\beq \label{eq:ais}
c_{(2,2)}=0\pm \frac{1}{\Lambda_{\mathrm{H}}^2}, \hspace{1cm}c_{(0,4)}=0\pm
\frac{1}{\Lambda_{\mathrm{H}}^4},
\eeq
where $\Lambda_{\mathrm{H}}$ is the typical mass of the resonance exchanged,
say $\Lambda_{\mathrm{H}}\sim 1$~GeV.
\section{Structure functions and azimuthal asymmetries} \label{sec:num}
In this section we will give the numerical results for the structure
functions integrated over the whole Dalitz plot, at fixed $Q^2$,
\beqn
w_{A,C,SA,SB,SC} &=& \int \de s_1 \, \de s_2 \,
W_{A,C,SA,SB,SC}(Q^2,s_1,s_2), \nonumber \\
w_{D,E,SD,SE} &=& \int \de s_1 \, \de s_2 \, {\mathrm{sign}}(s_1-s_2)\,
W_{D,E,SD,SE}(Q^2,s_1,s_2),
\eeqn
where the sign ordering is needed due to the Bose symmetry.
The limits of integration of $s_2$ for a given value of $s_1$ are $s_2^- <
s_2 < s_2^+$ where
\beq
s_2^{\pm} = 2 M_{\pi}^2 + 2 E_1 E_3 \pm 2 \sqrt{E_1^2 - M_{\pi}^2}
\sqrt{E_3^2 -  M_{\pi}^2} 
\eeq
and
\beq
E_1=\frac{Q^2 - M_{\pi}^2 - s_1}{2 \sqrt{s_1}}, \hspace{1cm}
E_3=\frac{1}{2}\sqrt{s_1}
\eeq
are the energies of the particles~1 and~3 in the 2--3 rest frame ({\em
i.e.}\ in the system where $\vec{p}_2 + \vec{p}_3 =0$),
while the bounds of $s_1$ are
\beq
4 M_{\pi}^2 < s_1 < \left( \sqrt{Q^2} - M_{\pi}^2 \right)^2.
\eeq
Due to  the presence of two identical particles an additional  factor $1/2$
has to be understood in the integration over the Dalitz plot.
\FIGURE{
\epsfxsize=12cm\epsfbox{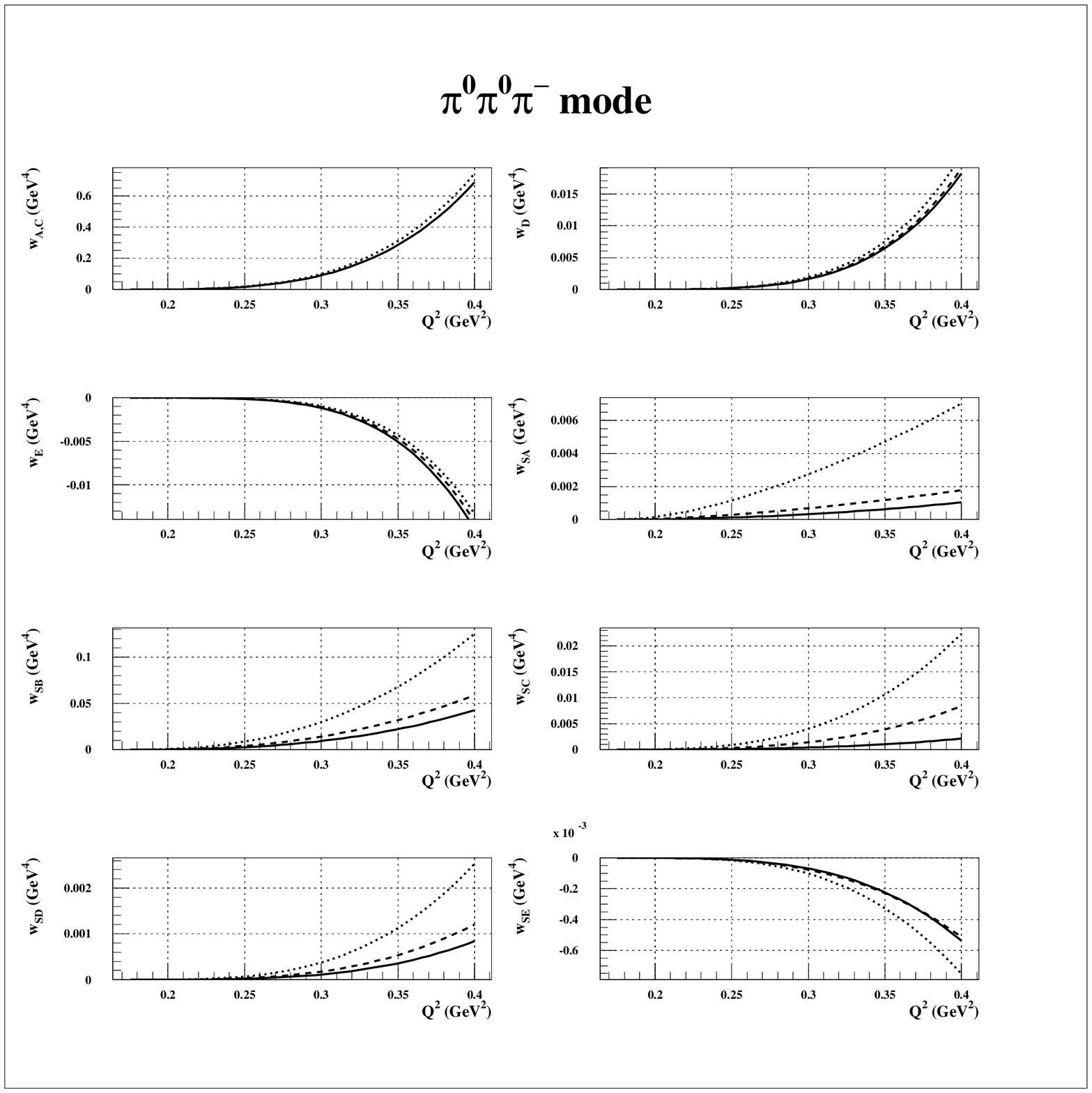}
   \caption{\label{fig:sf00-}\it
   Integrated structure functions for the $2 \pi^0 \pi^-$ charge mode.
   $w_A$ and $w_C$ (here and in the next figure) are indistinguishable at
   the scale of the plot. Solid, dashed and dotted lines refer
   to the three sets of values for $\alpha$ and $\beta$ of
   Eq.~(\ref{eq:trealfabeta}). They correspond respectively to the standard
   predictions, the central values extracted from the $K_{e4}$ experiment,
   and to the extreme case of tiny condensate.}  
    }
\FIGURE{
\epsfxsize=12cm\epsfbox{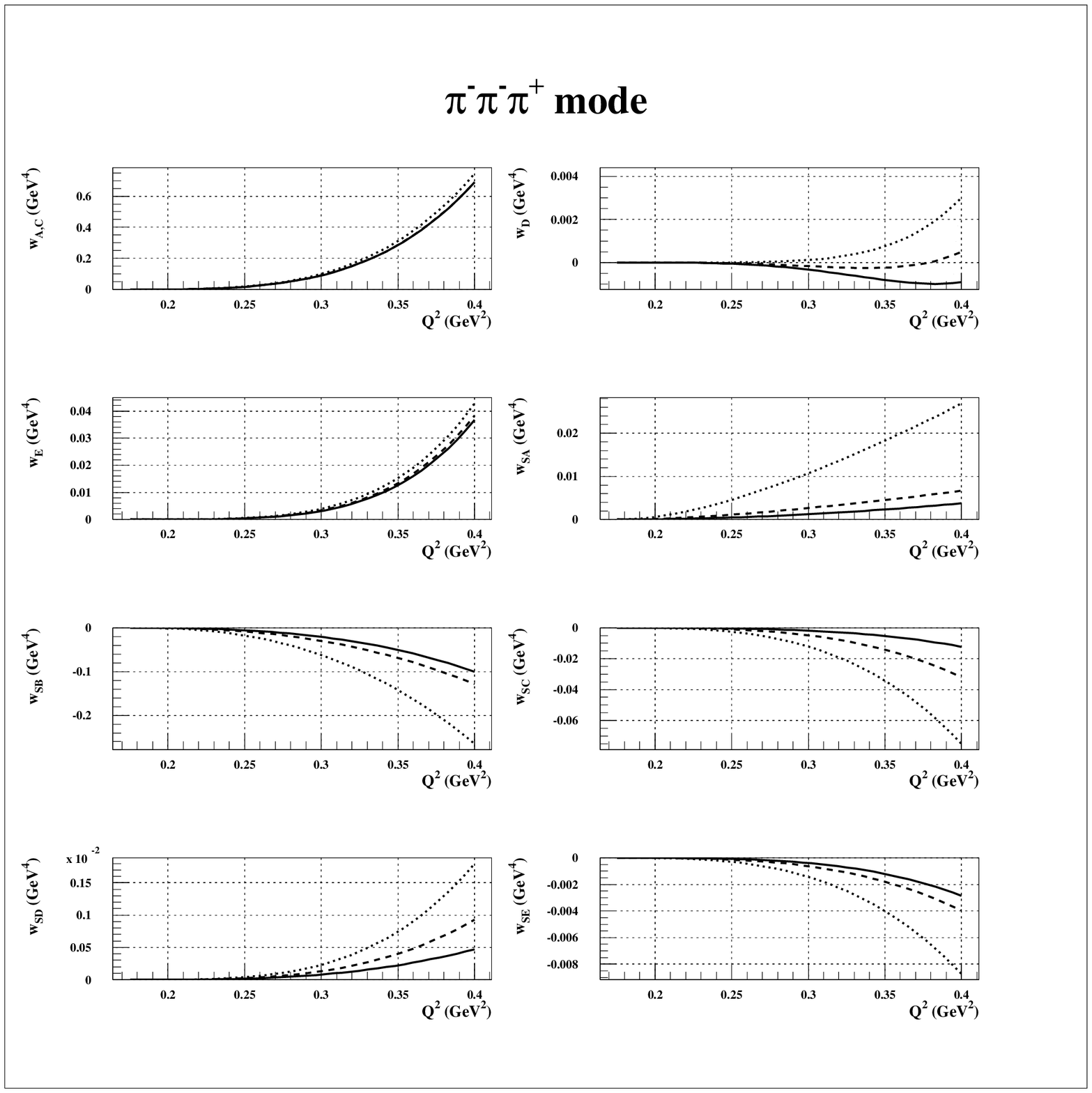}
   \caption{\label{fig:sf--+}\it
   Integrated structure functions for the $2 \pi^- \pi^+$ charge mode. The
   legend is as in the previous figure.}
    }
The results for both the charge modes are shown in
Figs.~\ref{fig:sf00-} and~\ref{fig:sf--+}. In both cases $w_A$ and $w_C$ are
practically indistinguishable at the scale of the plots. The reason for
that is, as already pointed out in Ref.~\cite{colangelo}, that
\beq
W_A-W_C=2 x_3^2 |F_1 - F_2|^2,
\eeq
where $F_1 - F_2$ is antisymmetric with respect to the exchange of the
particles~1 and~2, and, as it is clear from Eq.~(\ref{eq:H+-0}), the
antisymmetric part starts at order ${\cal O}(p^4)$.
$w_A$ is the main contribution to the total decay rate:
\beq
\de \Gamma = \frac{G_{\mathrm{F}}^2}{128 M_{\tau}} \frac{V^2_{\mathrm{ud}}}{(2 \pi)^5}
 \left[ \frac{ M_{\tau}^2 - Q^2}{Q^2} \right]^2
\frac{ M_{\tau}^2 + 2 Q^2}{3 M_{\tau}^2} W(Q^2) \de Q^2
\eeq
where
\beq \label{eq:wq2}
W(Q^2) = w_A + \frac{ 3 M_{\tau}^2}{M_{\tau}^2 + 2 Q^2} w_{SA}.
\eeq
and we can see that most of the other structure functions are much smaller
compared to it.
The solid, dashed and dotted  lines correspond to the three
sets of values  of Eq.~(\ref{eq:trealfabeta}) discussed above.  We see that
all the spin~1 structure functions practically do not depend on $\langle
\bar q q \rangle$ (except $w_D$ for the all charged mode, which, being zero
at the tree level, is very small and very much affected by the theoretical
uncertainties). 
On the contrary a rather strong dependence is observed not only in the
purely spin~0, but also in the interference terms $w_{SB},...,w_{SE}$. In
particular, among the latters, $w_{SB}$ is the largest one and its magnitude
is comparable to $w_A$. Clearly, $w_{SB}$ is the most appropriate structure
function where to look for the S-wave contribution.
From the  experimental  point of view, the sensitivity to the size of
$\langle \bar q q \rangle$ should therefore be expected in azimuthal angular
asymmetries (see Ref.~\cite{marbella}).  These are obtained by integrating
the differential decay rate~(\ref{eq:decayrate}) over all the variables
except $Q^2$ and the azimuthal angle $\gamma$, using the explicit angular
dependence of the functions $L_X$ found in Ref.~\cite{km}:
\beq
\de \Gamma = \frac{G_{\mathrm{F}}^2}{128 M_{\tau}} \frac{1}{(2 \pi)^5}
V^2_{\mathrm{ud}} \left[ \frac{ M_{\tau}^2 - Q^2}{Q^2} \right]^2
\frac{ M_{\tau}^2 + 2 Q^2}{3 M_{\tau}^2}\, f(Q^2,\gamma)\,W(Q^2) \de Q^2
\frac{\de \gamma}{2 \pi},
\eeq
with $W(Q^2)$  defined above  and the azimuthal distribution, normalized
to~1,
\beqn 
f(Q^2,\gamma) &=& 1 + \lambda_2 (Q^2,\beta_{\tau}) \left(
C^{\prime}_{\mathrm{LR}}\,\cos 2 \gamma + C^{\prime}_{\mathrm{UD}}\, \sin 2 \gamma
\right)\nonumber \\
&& + \lambda_1 (Q^2,\beta_{\tau}) \left( C_{\mathrm{LR}} \, \cos \gamma
+ C_{\mathrm{UD}}
\, \sin \gamma \right). \label{eq:azdistr}
\eeqn
The asymmetry coefficients $C^{\prime}_{\mathrm{LR}}$, $C^{\prime}_{\mathrm{UD}}$,
$C_{\mathrm{LR}}$ and $C_{\mathrm{UD}}$
in Eq.~(\ref{eq:azdistr}) are related to the Kuhn and Mirkes' structure
functions by the relations\footnote{
We are neglecting the  polarization of the $\tau$'s, which is  justified
 provided that the $Z$ exchange can be neglected  (far from the $Z$ peak).} 
\beqn
&&  C^{\prime}_{\mathrm{LR}}=\frac{1}{3} \left( 1 - \frac{Q^2}{M_{\tau}^2}
\right) \frac{3 M_{\tau}^2}{M_{\tau}^2 + 2 Q^2} \frac{w_C}{W},
\hspace{1cm}
 C^{\prime}_{\mathrm{UD}}=- \frac{1}{3} \left( 1 - \frac{Q^2}{M_{\tau}^2}
 \right)  \frac{3 M_{\tau}^2}{M_{\tau}^2 + 2 Q^2}  \frac{w_D}{W},
\nonumber \\
&& C_{\mathrm{LR}} = - \frac{\pi}{4}  \frac{3 M_{\tau}^2}{M_{\tau}^2 + 2
Q^2}  \frac{w_{SB}}{W}, \hspace{3cm}
C_{\mathrm{UD}} =
\frac{\pi}{4}  \frac{3 M_{\tau}^2}{M_{\tau}^2 + 2 Q^2}  \frac{w_{SD}}{W}.
\eeqn
The interesting quantity is the left-right asymmetry, whose coefficient
is related to $w_{SB}$.
The funtions $\lambda_i$ in Eq.~(\ref{eq:azdistr}) result from the
integration over the $\tau$-decay angle,
\beq
\lambda_i(Q^2,\beta_{\tau}) = \int_{-1}^1 \frac{ \de \cos \theta}{2}
P_i(\cos \psi),
\eeq
where $\beta_{\tau}$ is the $\tau$ velocity
\beq
\beta_{\tau} = \sqrt{1 - \frac{ M_{\tau}^2}{E_{\mathrm{beam}}^2}},
\eeq
and the angle $\psi$ is defined in Ref.~\cite{km}.
For  $\tau$'s produced at rest, which is the case for the $\tau$-charm
factories, this angle is $\cos \psi = 1$, so that the functions $\lambda_i$
are  equal to~1, whereas in the ultrarelativistic case, $\beta_{\tau}=1$,
they become
\beqn \label{eq:lambdaas}
\lambda_1(Q^2,1) &=& \frac{1}{\left( M_{\tau}^2 - Q^2 \right)^2} \left(
M_{\tau}^4 - Q^4 + 2 M_{\tau}^2 Q^2 \log \frac{Q^2}{M_{\tau}^2} \right),
\nonumber \\
\lambda_2(Q^2,1) &=& -2 + 3 \frac{M_{\tau}^2 + Q^2}{M_{\tau}^2 - Q^2}
\lambda_1 (Q^2,1).
\eeqn
\FIGURE{
 \centerline{\psfig{figure=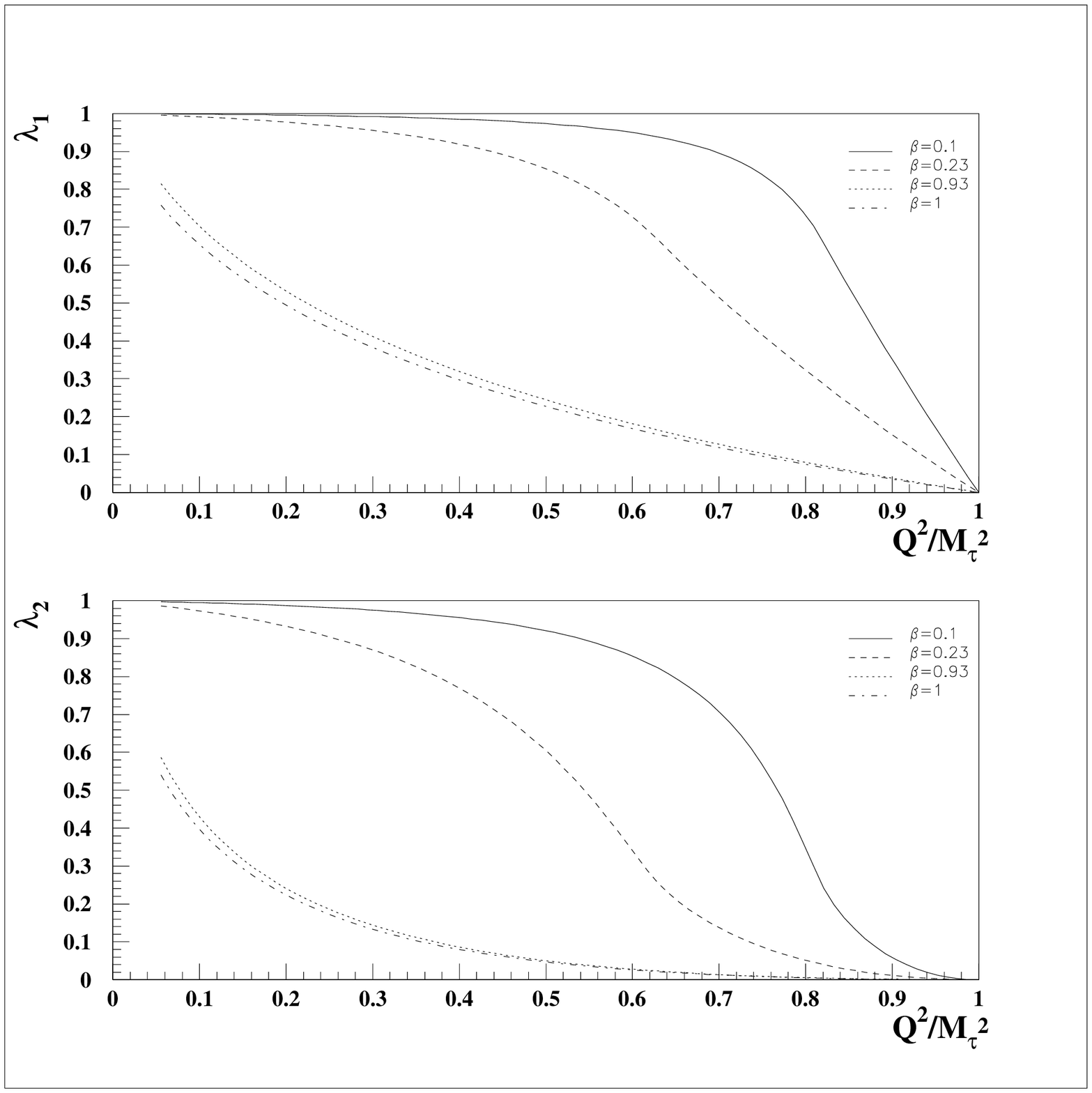,height=12cm,angle=0}}
\caption{\label{fig:lambdas}\it
   The functions $\lambda_1$ and $\lambda_2$ for different values of the
   $\tau$ velocity $\beta_{\tau}$.
    }
}
These two  functions are plotted for different values of $\beta_{\tau}$
in Fig.~\ref{fig:lambdas}. We see from the figure that, for high energy
machines, the functions $\lambda_i$ decrease the sensisivity in the
threshold region, but the effect is not so drastic, especially for
$\lambda_1$, which multiplies the coefficient $C_{\mathrm{LR}}$.
In order to take full advantage of the whole statistics we can consider the
integrated left-right asymmetry,
\beq \label{eq:alr}
A_{\mathrm{LR}}(Q^2) = \left| \frac{N_{\mathrm{R}}
-N_{\mathrm{L}}}{N_{\mathrm{R}}
+N_{\mathrm{L}}} \right|
\eeq  
where $N$ stands for the number of events from threshold up to a certain
$Q^2$ and the subscript L (R) refers to the events with $\gamma \in [0,\pi/2
] \cup [3\pi/2,2\pi]$  ($\gamma \in [ \pi/2 , 3 \pi/2]$).
This asymmetry\footnote{
The results for the asymmetries refer to the case of ultrarelativistic
$\tau$'s, where the asymptotic expressions of Eq.~(\ref{eq:lambdaas}) can be
used.} for both the charge mode is shown in
Fig.~{\ref{fig:alr}.
\FIGURE{
 \centerline{\psfig{figure=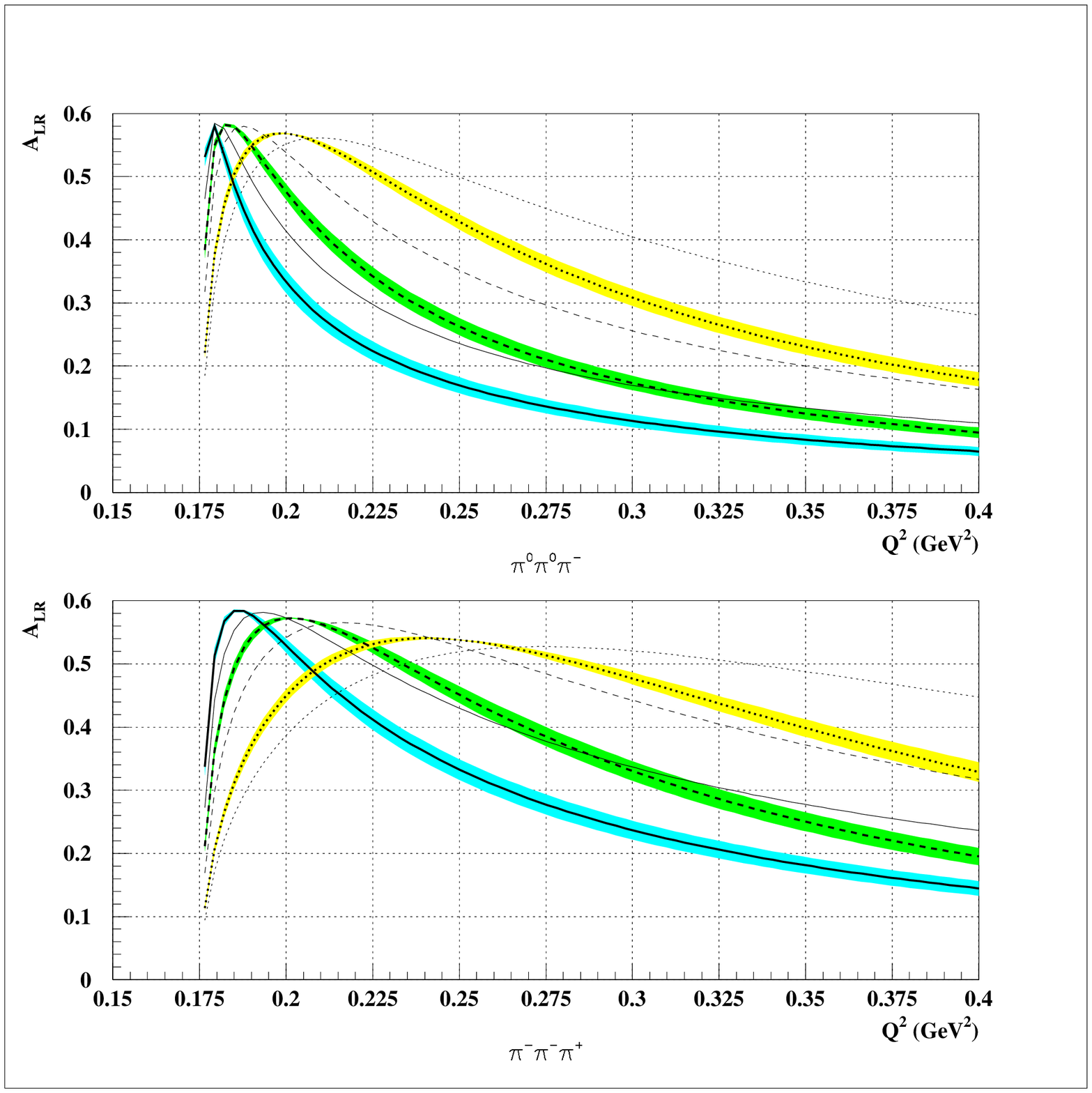,height=12cm,angle=0}}
\caption{\label{fig:alr}\it
   The left right asymmetry, for the two charge modes, as function of $Q^2$.
   The three curves, solid, dashed and dotted, inside the errorband
   correspond to the three different values of $\alpha$ and $\beta$ in
   Eq.~(\ref{eq:trealfabeta}) (like in
   Figs.~\ref{fig:sf00-} and~\ref{fig:sf--+}). The     corresponding lines
   outside the bands are the results at tree level.
    }
}
The solid lines correspond to the standard case,
the dashed lines  to the central ``experimental'' values of
Eq.~(\ref{eq:alfabetaexp}) and the dotted ones to the extreme case of tiny
condensate [Eq.~(\ref{eq:trealfabeta})]. The bands represent the theoretical
uncertainties, coming from the errors on the low energy constants
[Eqs.~(\ref{eq:lis}),~(\ref{eq:l6b}) and~(\ref{eq:ais})\footnote{
In the estimation of the error from Eq.~(\ref{eq:ais}) we have taken $\hat m
\sim 20$~MeV.}]  and from a
variation of the $\chi$PT renormalization scale $\mu$, at which the estimate
(\ref{eq:ais}) is supposed to hold, between 500~MeV and 1~GeV.
Particular attention has been payed to the sensitivity of the structure
functions to the two constants $\bar l_1$ and $\bar l_2$. One sees from
Table~\ref{tab:sens} that for all the asymmetries considered in this paper
the uncertainty coming from these two constants is well under control. This
would not be the case for the up-down asymmetries, related to the structure
functions $w_D$ and $w_{SD}$.
\TABLE{
\begin{tabular}{c|c|c|c|c|c|c|c|c|c}
$X$ & $A$ & $C$ & $D$ & $E$ & $SA$ & $SB$ & $SC$ & $SD$ & $SE$ \\
\hline
$\frac{1}{w_X} \frac{ \partial w_X^{00-}}{\partial \bar l_1}$ &
0.02 & 0.02 & 0.01 & 0.02 & -0.03 & -0.004 & 0.04 & -0.02 & -0.03 \\
$\frac{1}{w_X} \frac{ \partial w_X^{00-}}{\partial \bar l_2}$ &
-0.02 & -0.02 & 0.3 & -0.04 & -0.06 & -0.05 & 0.009 & 0.2 & -0.2 \\
\hline
$\frac{1}{w_X} \frac{ \partial w_X^{--+}}{\partial \bar l_1}$ &
0.02 & 0.02 & 8 & 0.005 & -0.03 & -0.007 & 0.02 & 0.6 & -0.05 \\
$\frac{1}{w_X} \frac{ \partial w_X^{--+}}{\partial \bar l_2}$ &
-0.02 & -0.02 & 8 & -0.02 & -0.07 & -0.05 & -0.006 & 0.6 & -0.07 
\end{tabular}
\caption{\it The dependence of the nine integrated structure functions $w_X$
on $\bar l_1$ and $\bar l_2$. The derivatives are computed at
$Q^2=0.35{\mathrm{GeV}}^2$.}
\label{tab:sens}
}
The lines outside the bands in Fig.~\ref{fig:alr} correspond to the result
at tree level.
We see that the one-loop corrections to the asymmetry are rather important,
not exceeding 
however 30~\%. Assuming a geometrical behavior of this series, the two-loop
effects are expected to remain in the range of 10~\%, which at present stage
should be considered as an additional systematic error to the one shown in
Figs.~\ref{fig:alr} and~\ref{fig:a12lr}. Notice however that a better control of
this systematic error is conceivable: $i)$~most of this uncertainty  comes
from the spin~1 part of the matrix element (see also Fig.~\ref{fig:br}), and
consequently an inclusion of the resonance contribution which dominates in
this channel might shed more light on the problem; $ii)$~the strong final
state interaction in the S-wave can be treated more precisely by the
standard procedure of unitarization (cfr. Ref.~\cite{kl4}).

In the case of $\tau$ leptons produced close to threshold  ($\tau$-charm
factories), the above asymmetries are somewhat larger since in this case the
kinematical functions $\lambda_1=\lambda_2=1$ increase the contribution of
the low-$Q^2$ region to the integrated asymmetries.
Taking as reference the value $Q^2 = (600$~MeV$)^2$,
$A_{\mathrm{LR}}$ for the all charged mode goes from $(60 \pm 6)\%$, for
tiny condensate ($\alpha = 4$), to $(28\pm 4)\%$ for the standard case.      
We emphasize once more that the reason for such
a big symmetry breaking effect of the S-wave is the kinematical suppression
of the P-wave near 
the threshold. Another asymmetry where the same argument applies is
$A^{\prime}_{\mathrm{LR}}$, plotted in Fig.~\ref{fig:a12lr} (the legend is
the same as in the previous figure), which corresponds to the
coefficient $C^{\prime}_{\mathrm{LR}}$. $A^{\prime}_{\mathrm{LR}}$ is defined
analogously to Eq.~(\ref{eq:alr}) with L
referring to the events with $\gamma \in [\pi/4,3 \pi/4] \cup [5 \pi/4,
7\pi/4]$  and R to the complementary interval.
Eventhough this asymmetry is of less practical utility than
$A_{\mathrm{LR}}$, it is remarkable that
the separation of the three cases near the threshold is entirely due to the
purely spin~0 structure function, which is suppressed by two powers of the
quark mass.
\FIGURE{
 \centerline{\psfig{figure=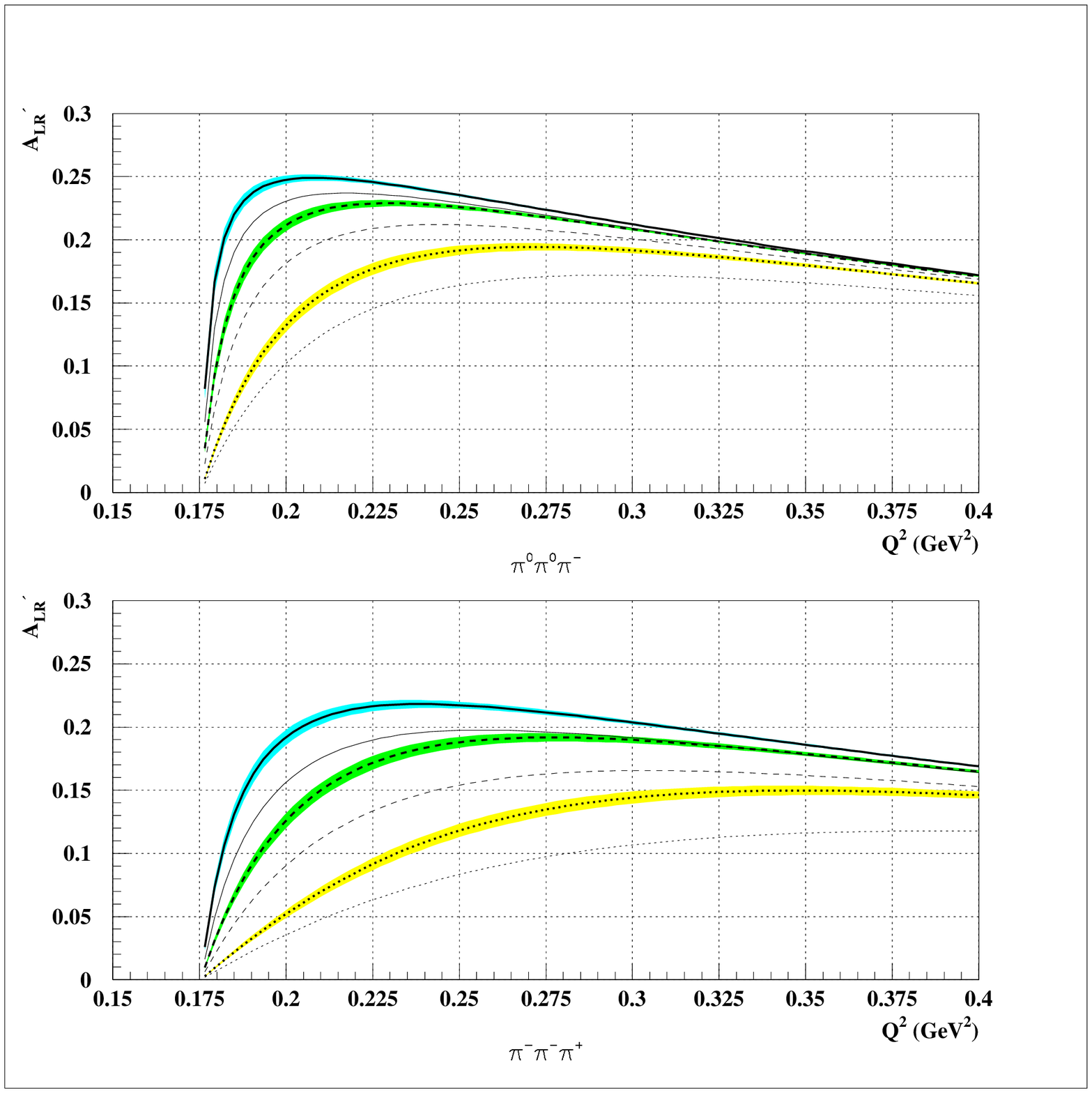,height=12cm,angle=0}}
\caption{\label{fig:a12lr}\it
   The  asymmetry $A^{\prime}_{\mathrm{LR}}$ for the two charge modes, as
   function of $Q^2$. The legend is like in Fig.~\ref{fig:alr}.
     }
}
Now that we have the predictions for the azimuthal asymmetries, it is useful
to know how many events are expected in the near-threshold region. Using the
experimental value of the $\tau$ lifetime, $\chi$PT is able to predict the
integrated branching ratio, that is the number of events from threshold up
to  a certain $Q^2$, divided by the total number of $\tau$ produced.
This quantity is plotted in Fig.~\ref{fig:br}.
It turns out that it is very similar for both the charge modes.
The uncertainties  coming from the low energy constants and from a variation
of the $\chi$PT renormalization scale between 500~MeV and 1~GeV are
 invisible at the scale of the plot. On the other hand, we see that the
 convergence of the $\chi$PT series is much worse than in the case of the
 asymmetries, as already pointed out (typically the tree level contribution
 is half  of the total one-loop result; cfr. the discussion in
 Ref.~\cite{colangelo}). We thus conclude that, in contrast to the
 asymmetries, the one-loop $\chi$PT predictions for the branching ratio are
 less reliable, and should not be used as a basis for a precision test of
$\chi$PT.
\FIGURE{
 \centerline{\psfig{figure=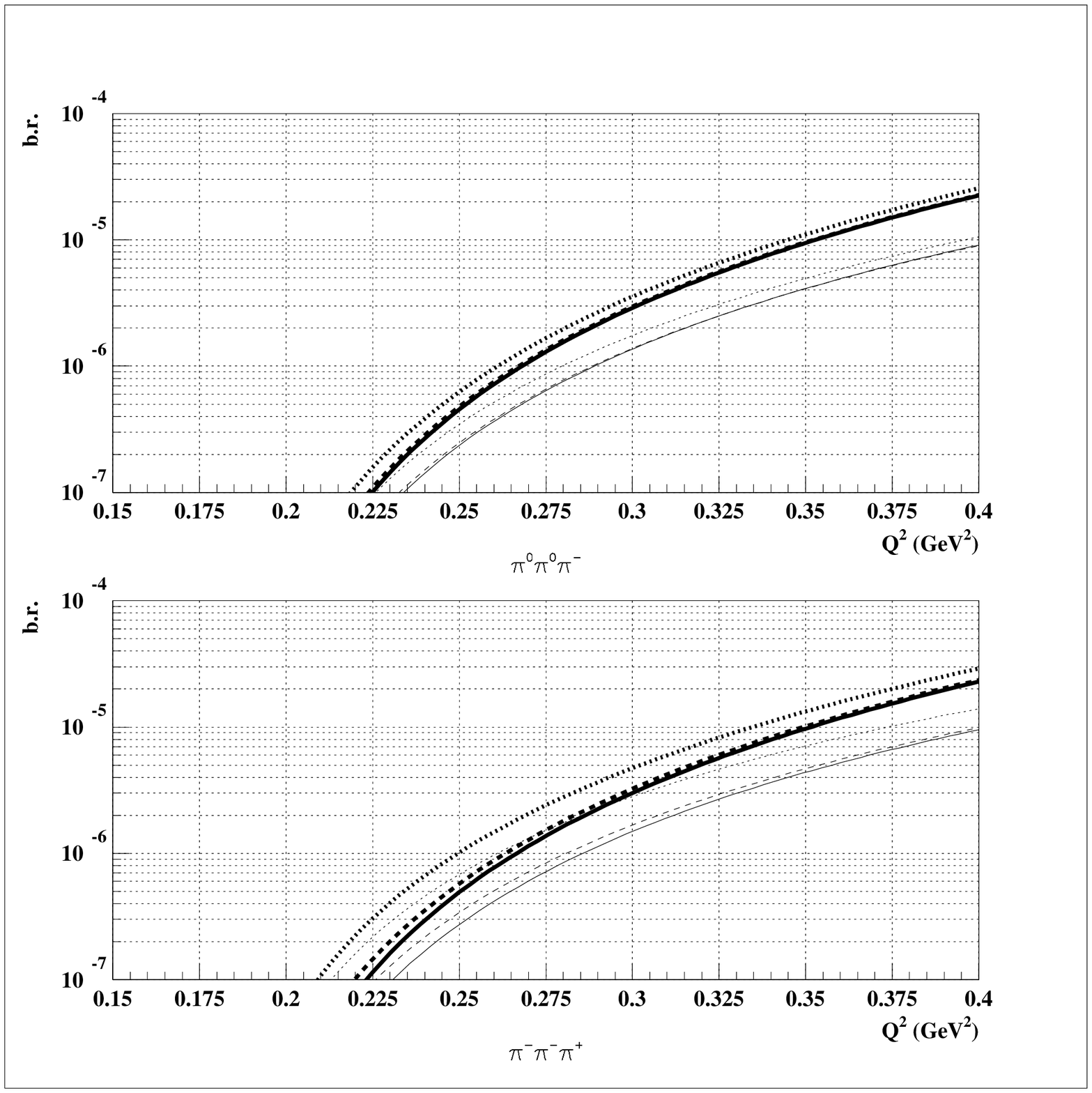,height=12cm,angle=0}}
\caption{\label{fig:br}\it
   The integrated branching ratio for the two charge modes. The solid,
   dashed and dotted lines correspond respectively to the three values of
   Eq.~(\ref{eq:trealfabeta})
   for $\alpha$ and $\beta$. The thicker lines are the one loop results and
   the thinner the corresponding tree level ones.}
}
With this in mind, according to Fig.~\ref{fig:br},  the number of
events with $Q^2 < (600$~ MeV$)^2$ is expected to be of the order of $10^{-5}$
times the number of $\tau$'s produced.
On the experimental side, the largest statistics, up to now, has been collected
by CLEO, with a total number of $\tau$ pairs of $\sim
10^7$, that is a hundred of events below $($600 MeV$)^2$ for each charge
mode of the decays $\tau \rightarrow 3 \pi \nu_{\tau}$.
With such statistics, measuring  $A_{\mathrm{LR}}$  for the all charged
mode, one should already be able to extract $\alpha$, and  then the quark
condensate, with a precision comparable to the present one, summarized in
Eq.~(\ref{eq:alfabetaexp}).
According to M.~Perl (see Ref.~\cite{perl}),  a factor~$\sim 5$ increase in
statistics  can reasonably be expected in the near future, thanks to the
new high luminosity $B$-factories.
That could allow a much sharper  measurement of  $\alpha$, thus providing an
interesting and completely independent cross-check of future precise
determinations from $\pi\pi$ scattering. 
\section{The size of the spin~0 spectral function} \label{sec:qmas}
The scalar structure function $w_{SA}(Q^2)$ is proportional to the
three-pion component of the spin~0 $\tau$ spectral function, that is the
spectral function of the divergence of the axial current. This quantity
enters in the sum rule determinations of the light quark mass $\hat m$. The
two-point function relevant for such determinations is
\beq
\psi_5( Q^2) = i \int \de^4 x {\mathrm{e}}^{i Q\cdot x} \langle 0 | T
\left\{ \partial^{\mu} \left( \bar u \gamma_{\mu} \gamma_5 d \right) (x) \partial^{\nu}
\left( \bar d \gamma_{\nu} \gamma_5 u \right) (0) \right\} | 0 \rangle,
\eeq
whose imaginary part is proportional to the spectral function
\beq
\frac{1}{\pi} {\mathrm{Im}} \psi_5 (Q^2) = \rho (Q^2) = \frac{1}{2 \pi}  \sum_n  ( 2 \pi
)^4 \delta^4 (Q - p_n ) \left| \langle n | \partial^{\mu} \left( \bar d
\gamma_{\mu} \gamma_5 u \right) | 0 \rangle \right|^2.
\eeq
Using the Cauchy's theorem and the analytic properties of the two-point
function, one can relate the integral of the spectral function along the
positive $x$-axis, to a contour integral of $\psi_5(Q^2)$ at high energy.
Evaluating the latter with the help of perturbative QCD together with the
OPE technique for the non perturbative contributions, one arrives to the
class of sum rules 
\beq \label{eq:cauchy}
\int_0^{s_0} \de Q^2 Q^{2 n} {\mathrm{Im}} \psi_5 (Q^2) = - \frac{1}{2 \pi
i} \oint_{|Q^2 | = s_0} \de Q^2 Q^{2 n} \psi_5 (Q^2).
\eeq
The three-pion component  of the spin~0 spectral function reads
\beq
\rho_{3\pi}(Q^2) = \frac{1}{256 \pi^4} \left[ w_{SA}^{00-} (Q^2) +
w_{SA}^{--+} (Q^2) \right].
\eeq  
This is the hadronic phenomenological input that one has to insert, in the
low energy part of the
l.h.s. of Eq.~(\ref{eq:cauchy}) in order to evaluate the sum rule. Due to the
smallness of the $w_{SA}$ structure
functions, a direct experimental information on $\rho_{3\pi}$ is lacking.
Therefore a model for the three-pion continuum has to be used, which should
include the observed resonances in this channel. Yet the overall
normalization of the  spectral function modelized in this way, remains
unknown. It has been proposed in Refs.~\cite{dder,bpder} to normalize the
hadronic model to the low energy behavior predicted by standard $\chi$PT.
However, as we have seen in the previous section, the low energy behavior of
$w_{SA}$ depends very much on the size of the quark condensate. 
  \FIGURE{
 \centerline{\psfig{figure=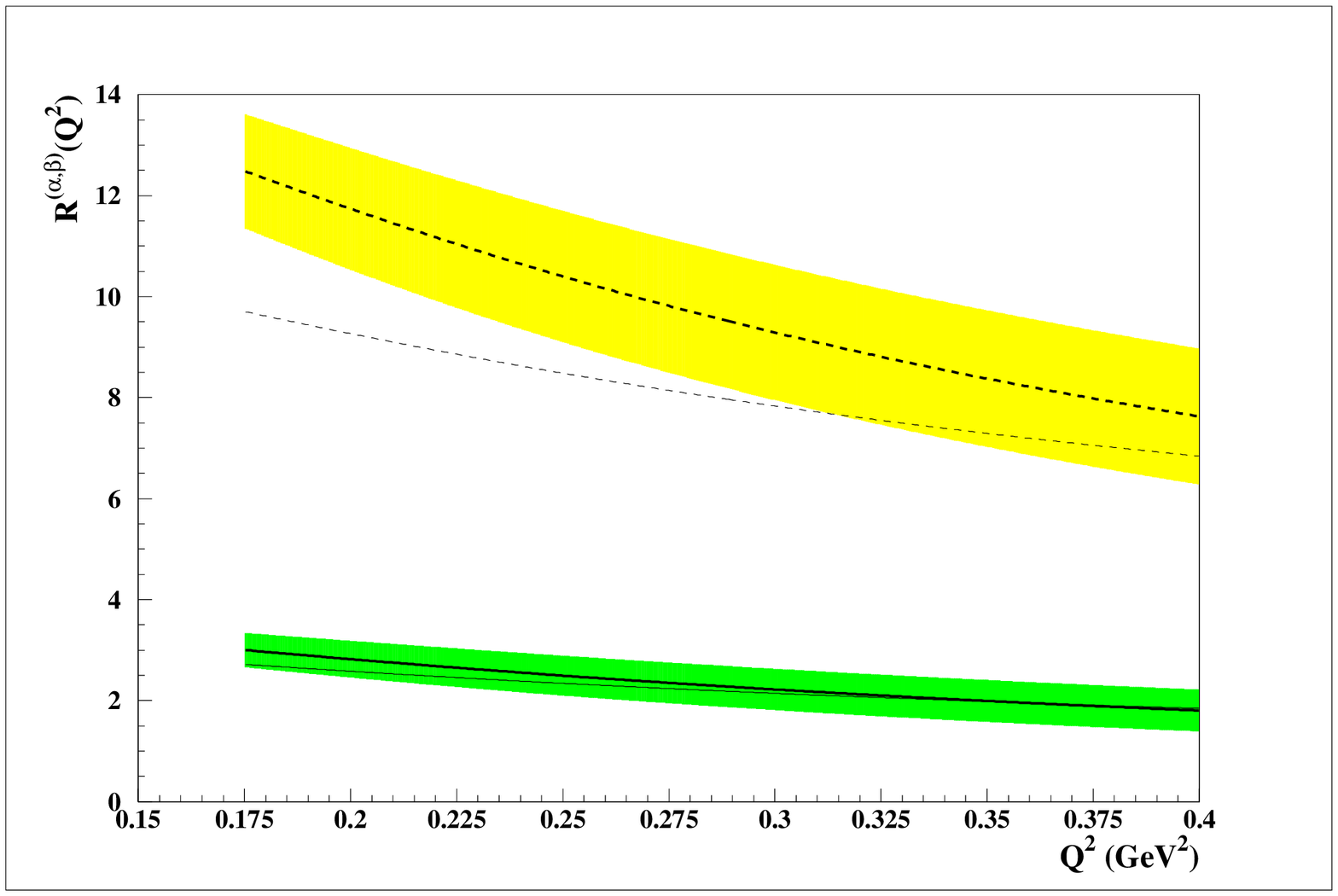,height=9cm,angle=0}}
\caption{\label{fig:renf}\it
	The three-pion component of the spectral function  $\rho_{3
	\pi}(Q^2)$ for $\alpha=2.16$, 	$\beta=1.074$ (solid line) and for
	$\alpha =4$, $\beta = 1.16$ (dashed line), normalized to the
	standard $\chi$PT predictions. The lines outside the errorbands are
	the tree level results.
}
}
In Fig.~\ref{fig:renf} we plot the ratio 
\beq
R^{(\alpha,\beta)}(Q^2) = \frac{\rho^{(\alpha,\beta)}_{3\pi}(Q^2)}{\rho_{3
\pi}^{\mathrm{st}}(Q^2)} 
\eeq
of  the 3-pion component of the spectral function for a given value of
$\alpha$ and $\beta$ divided by the standard $\chi$PT prediction.
The two bands correspond to the second and third set of values of
Eq.~(\ref{eq:trealfabeta}); also shown are the tree level results.
We see that the low energy spectral function increases by a factor of $\sim
12$ in  extreme case of $\alpha =4$, and the output of the sum rule for
$\hat m^2$ can be expected to increase accordingly. In Ref.~\cite{marbella}
the threshold behavior of the same  ratio has been analyzed neglecting
$M_{\pi}^2$ in the integration over the Dalitz plot. The latter
approximation is actually not very good near the threshold and 
overestimates the result: at tree level, in the extreme case of $\alpha =4$,
the ratio $R^{(\alpha,\beta)}(9 M_{\pi}^2)$ was found to be~13.5. However,
this effect is compensated by the increase due to the one-loop corrections.
\section{Concluding remarks} \label{sec:conc}
The understanding of the mechanism of DBCHS in QCD requires the
knowledge of the size of the quark anti-quark condensate. G$\chi$PT is the
appropriate tool to pin down this quantity from low energy experiments.
The physical observable that will presumably allow to do that is the low
energy $\pi\pi$ scattering.
In this paper the SU(2)$\times$SU(2) generalized chiral lagrangian is
constructed to ${\cal O}(p^4)$ and renormalized at one-loop level.
We have presented an alternative way of extracting $\langle \bar q q
\rangle$, from the $\tau$ decays into three pions, performing an analysis at
one-loop level of G$\chi$PT. The dependence on the
condensate is contained in the S-wave of these decays, which is an effect of
explicit chiral symmetry breaking.
However, due to a kinematical suppression of the P-wave, compared to the
S-wave, the latter shows up as a detectable effect near the threshold,
through the azimuthal left-right asymmetries.
At $Q^2$ large enough to allow for an experimental determination ($Q^2 \sim
0.35 {{\mathrm{GeV}}}^2$) it turns out that the integrated left-right
asymmetry is larger for the all charged mode (see Fig.~\ref{fig:alr}). In
the case of ultrarelativistic $\tau$'s (relevant, {\em e. g.}\, , for CLEO)
the predictions of G$\chi$PT for this asymmetry, integrated from threshold
up to $Q^2 =0.35 {{\mathrm{GeV}}}^2$, range from $(17 \pm 2 )$ \% (in the
standard case) up to $(40 \pm 2)$ \%, depending on the size of $\langle \bar
q q \rangle$. These errorbars do not take into account the uncertainty
coming from the higher chiral orders. We estimate the latter to be $\sim 10
\%$ of the total result. Therefore a measurement significantly larger than
$20 \%$ would already signal a departure from the standard picture of large
condensate. Unfortunately the number of events expected in this region is
rather small: the branching ratio integrated from threshold to  $Q^2 =0.35
{{\mathrm{GeV}}}^2$  is of the order of $10^{-5}$ for each charge mode.
The presently available  statistics ($10^7$ $\tau^+\tau^-$ pairs of CLEO)
will allow to measure this asymmetry, but not to distinguish between the
standard and the extreme generalized case.
However improvements can be expected from both theoretical and experimental
sides.
The theoretical predictions can in principle be sharpened by an analysis
``beyond one loop'', which should include, in the spin~1 form factor, the
effect of the resonance $a_1$ (cfr. Ref.~\cite{paco}) and, in the spin~0 form factor, the strong
final state interactions of the S-wave (unitarization procedure \cite{kl4}). From the
experimental point of view an improvement in statistics by a factor $ 5 \div
10$ seem reachable. That would allow a real measurement af the parameter
$\alpha$, related to $\langle \bar q q \rangle$, providing a completely
independent cross-check of future precise determinations from $\pi\pi$
scattering.

\acknowledgments
It is a pleasure to thank Marc Knecht for correspondence and
discussions.  L.G. wishes also to acknowledge useful discussions with Toni
Pich.
We would like to thank Dominique Vautherin for
the hospitality at LPTPE, Universit\'e Pierre et Marie Curie, Paris~VI,
where part of this work has been done.
This work has been partially supported by the EEC-TMR Program, Contract N.
CT98-0169 (EURODA$\Phi$NE).

\end{document}